\documentclass[final,5p,times,twocolumn,authoryear]{elsarticle}

\usepackage{amssymb}

\usepackage{lineno}

\usepackage{subcaption}

\journal{Journal of Volcanology and Geothermal Research}

\usepackage[usenames,dvipsnames]{color}

\begin{document}

\begin{frontmatter}



\title{Effects of magma-induced stress within a cellular automaton model of volcanism}


\author[nu]{Olivia J.\ Butters}

\author[nu]{Graeme R.\ Sarson\corref{cor1}}
\cortext[cor1]{Corresponding author.}
\ead{g.r.sarson@newcastle.ac.uk}

\author[nu]{Paul J.\ Bushby}

\address[nu]{School of Mathematics \& Statistics, Newcastle University, Newcastle upon Tyne NE1 7RU, UK}

\begin{abstract}
The cellular automaton model of 
Piegari, Di Maio, Scandone and Milano,
{\it J.\ Volc.\ Geoth.\ Res.}, {\bf 202}, 22-28 (2011)
is extended to include magma-induced stress
(i.e.\ a local magma-related augmentation of the stress field).
This constitutes a nonlinear coupling between the magma and stress fields
considered by this model, 
which affects the statistical distributions of eruptions obtained.
The extended model retains a power law relation 
between eruption size and frequency for most events,
as expected from the self-organised criticality inspiring this model;
but the power law now applies for a reduced range of size,
and there are new peaks of relatively more frequent eruptions
of intermediate and large size.
The cumulative frequency of repose time between events 
remains well modelled by a stretched exponential function of repose time
(approaching a pure exponential distribution only for the longest repose times),
but the time scales of this behaviour are slightly longer,
reflecting the increased preference for larger events.
The eruptions are relatively more likely to have high volatile (water) content,
so would generally be more explosive.
The new model also naturally favours a central `axial' transport conduit,
as found in many volcano systems,
but which otherwise must be artificially imposed within such models.
\end{abstract}


\end{frontmatter}


\section{Introduction}
\label{sect:intro}


Volcanism occurs in a variety of styles, ranging from effusive to explosive, with orders of magnitude variation in the volume of ejecta and in the repose time between eruptions. This reflects the wide range of tectonic settings, magma compositions, and variations in structure of the volcanic conduit(s) through which magma is transported \citep{NewhallToG,SiebertBook}. Despite these variations between different systems, some statistical features of eruptions are well-established. Quantifying the magnitude of an eruption using the  Volcanic Explosivity Index, VEI \citep[][]{Newhall1982}, \cite{Simkin1993} demonstrated an exponential relation between the frequency of Holocene eruptions and their magnitude, for VEI values in the range 2--7. 
This corresponds to a power law relation between eruption frequency
and the volume of ejecta,
since a unit increase in VEI corresponds to a factor of 10 increase
in ejecta.
This power law relation can be compared to the well-known Gutenberg--Richter law for earthquakes \citep{Gutenberg1956}, with VEI playing a similar role to the earthquake magnitude $M$. Indeed, many parallels can be drawn between volcanic activity and tectonic activity on fault zones \citep{NewhallToG}. 

\par Power law behaviour is often a signature of Self-Organised Criticality, SOC \citep[e.g.][]{JensenBook}. Ideas of self-organisation in volcano fracture systems go back at least to \cite{Shaw1991}, who considered fractal percolation networks beneath Hawaiian volcanoes, and the nonlinear dynamics linking these to tremor processes. Additional support for SOC in volcanism comes from the `pink' (i.e.\ self-similar, $1/f$) spectrum of noise observed in long period volcanic seismicity \citep{Lachowycz2013}. Such a spectrum is often associated with SOC processes. 

\par Cellular automata (CA) often exhibit SOC, and such systems have long been used to model seismicity.  \cite{Bak1989} extended their original sandpile model \citep{Bak1988} to consider earthquakes, and noted the connection with the Gutenberg--Richter law. Olami, Feder and Christensen (1992; OFC) made such CA models non-conservative, to allow for the energy loss in seismic motion, and the OFC earthquake model has become one of the `standard' SOC systems \cite[e.g.][]{JensenBook}. The OFC model is an abstract representation of the `slider block' model of seismicity \citep{Burridge1967}, with the equations of motion of the spring blocks being replaced by a CA consisting of a grid of cells on which the associated stress distribution evolves according to a simple set of rules \cite[e.g.][]{TurcotteBook}. 

\par The 2D OFC model is normally assumed to represent an abstract section of a fault plane
subject to a constant applied stress;
the earthquake events arise from local `stick-slip' behaviour,
which might remain localised (small events) or trigger an `avalanche'
of events involving neighbouring cells (leading to larger events).
\cite{Piegari2008} instead applied the model
to a vertical section of crust below a volcano, 
with the constant rate of stress representing a combination 
of the regional stress field and the local stresses associated with the volcanism.
To this they added a linked CA model, representing the presence of magma 
within the volcano, with rules
for the feeding of the system from an underlying reservoir, 
for the movement of magma under buoyancy, 
and for the eruption of magma as lava at the surface.
Within this model, magma movement is only allowed within the network of
fractured cells created by earthquake events;
this model therefore embodies the `magma batch' mechanism advanced by 
\cite{Scandone2007}, appropriate for closed conduit volcanoes,
and presented as a model for the late 20th century activity of
Mount St Helens (1980--2004) and Pinatubo (1991).

The original model of \cite{Piegari2008} was extended 
in subsequent papers,  
to model the volatile components within the magma \citep{Piegari2011},
and to investigate different background density profiles \citep{Piegari2012}
and the effect of a low-density surface layer \citep{Piegari2013};
some of the details of these models are presented and discussed
in sections~\ref{sect:model} and \ref{sect:discuss}. 
While these models are highly idealised, 
they produce eruptions ranging greatly in size 
(i.e.\ the number of cells of magma involved in the eruption), with the expected power law relation between frequency and size holding across most of the range. The system thus appears to be a useful working model 
for the style of volcanism described by \cite{Scandone2007}.

While the papers of Piegari et al.\ usefully extend the CA model of earthquakes to volcanism, they do not allow for any feedback from the magma activity upon the local seismicity (potentially an important effect within linked seismic-volcanic systems). Many studies corroborate the association between seismicity and magma activity. 
The occurrence of volcano-tectonic (VT) events --- 
`normal' tectonic earthquakes due to brittle fracture (shear failure), 
swarms of which may occur as precursors of eruptions ---
is often explained via changes in the stress field directly caused by the rising magma,
and by later relaxation \citep[e.g.][]{NewhallToG}.
The fracturing of the country rock by a propagating crack of buoyant magma
has often been studied as a fluid mechanical problem
\citep[e.g.][]{Emerman1986,Lister1991}.
\cite{Kilburn1998} note that direct magmatic stresses 
need not be the only effect,
and propose a model for subcritical rock failure
due to progressive weakening, most likely due to stress corrosion:
a stress-enhanced chemical reaction due to circulating fluids.
Such circulation may nevertheless
also be associated with nearby magma intrusion. VT events contrast with long-period (LP; low-frequency) events associated with the degassing of magma at shallow depths, and with dome formation \citep{Neuberg2000}.
Non-explosive magma fragmentation, due to degassing,
may also lead to the creation of intermittent fracture networks 
near the magma conduits \citep{Gonnermann2003}.

Although different in nature, both VT and LP events could be modelled 
within the CA system by feedback from the magma field to earthquake events.
We do not propose any specific mechanism for magma-induced stress, 
but instead (in the idealised spirit of the CA model) simply introduce a local enhancement of the strain rate in the vicinity of
magma, as described in detail in section~\ref{sect:model}.
As a final comment, we note that \cite{Scandone2007} suggest that
the complexity of the fracture system may systematically increase with time
during episodes of volcanism;
this possibility is absent from the models
of Piegari et al.\ (where the stress field is independent of the magma),
but not for our model (where increased magma in the system will naturally
lead to more widespread fracture networks).


\section{Model}
\label{sect:model}

{In this section, we summarise the details of the model, 
which builds on that of Piegari et al.\ (2008, 2011). 
After describing the essential features of the Piegari et al.\ model ---
the fracture, magma movement and degassing algorithms --- 
we move on to describe the crucial new feature introduced:
the augmentation of the local stress in the vicinity of magma.}


\subsection{{Earlier model: Piegari et al.\ (2008, 2011)}}

\subsubsection{{Fracture model}}
\label{sect:OFC}

As introduced above, the CA model for volcanism is based on
the OFC earthquake model \citep{Olami1992},
using it to simulate fractures in the country rock 
caused by volcanic tectonic activity.
The OFC model defines a stress field $f_{i,j}$,
on a 2D grid of size $L \times L$, 
with $1 \leq i \leq L$, $1 \leq j \leq L$.
In {the} volcanic context, 
this corresponds to a vertical subsection beneath
a volcano, with $i$ labelling the vertical axis
(with $i=L$ the row at the bottom of the grid),
and $j$ the horizontal axis.
Each cell is initially assigned a random stress value $f_{i,j}$, 
uniformly distributed in the range $0  \leq f_{i,j} < 1$.
In the homogeneous, isotropic case considered {in earlier work},
the stress at every cell is increased at constant strain rate $\nu$.
Evolving the system via time-steps of size $\Delta t$,
the stress at each cell varies as
$f_{i,j} \longrightarrow f_{i,j} + \nu\Delta t$.
This behaviour continues until the stress in any cell 
reaches the critical value, $f_{\rm crit}$. This critical value is the same for all cells and, working with nondimensional variables, we can take $f_{\rm crit}=1$.
When a cell reaches this threshold value it fractures:
a proportion of the stress, determined by a model parameter $\epsilon$,
is distributed to the neighbouring cells, 
and the stress on the original cell is reduced to zero. 
Since each cell has a maximum of 4 neighbours (in the interior of the grid), 
the range $0 \le \epsilon \le 0.25$ is possible,
with $\epsilon=0.25$ corresponding to a conservative system.
{Following \cite{Piegari2011},
we here consider a non-conservative system with $\epsilon = 0.2$.}
This redistribution of stress may cause the neighbouring cells 
to reach the critical value and fracture,
resulting in further redistributions of stress;
this process repeats, in an `avalanche' effect, 
until the system is everywhere stable again
(i.e.\ no cells remain with $f_{i,j}\ge f_{\rm crit}$). 
These repeated stress relaxations are considered to occur instantaneously,
before the next global strain increment, 
and constitute a single earthquake event. 
The net effect is to produce a network of fractured cells, 
involving between 1 and $L^2$ cells
(with the latter, system-wide events being vanishingly rare),
through which magma may move.
The probability of fracture events involving $N$ cells 
follows a power law in $N$,
corresponding to the Gutenberg--Richter law.

\par Since the stress-loading process between fracture events 
is constant and homogeneous, 
the time until the next event can be calculated by observing the
maximum value of $f_{i,j}$ after the preceding event.
The next event will happen when the stress has increased by
$\delta f = 1 - {\rm max}(f_{i,j})$.
With a constant stress-rate $\nu$, this will happen in time
$\delta t=\delta f/\nu$.
Rather than incrementing time in steps of $\Delta t$, 
we therefore step ahead to the next event by incrementing the time by
$\delta t$, and increasing the stress on all cells by $\delta f$. 
This is equivalent to time-stepping in the limit
$\Delta t\longrightarrow 0$, but is computationally more efficient.
We work with non-dimensional time units, 
defined with respect to the stress-rate $\nu$.
In practice our calculations therefore use $\nu=1$;
if dimensional times are required, they could be calculated by multiplying
our non-dimensional values by the time unit $\tau=f_{\rm crit}/\nu$, 
where $f_{\rm crit}$ and $\nu$ 
are the physical values of interest.

The OFC model has been studied with various forms of boundary conditions
at the edges of the grid. 
While this leads to quantitatively different results, 
the qualitative behaviour is unaffected. 
Here we use `rigid' boundary conditions
(as if the cells at the boundaries were connected to a rigid frame 
moving with the driving plate, 
within the slider-block model motivation for the OFC model). 
This is implemented using a grid of size $(L+2) \times (L+2)$,
with the stress at the (`dummy') external cells ($i,j=0,L+1$) 
being reset to zero after each event.
These conditions are non-conservative, with an additional loss of stress 
when boundary cells fracture:
$\epsilon f_{i,j}/{2}$ at corner cells,
$\epsilon f_{i,j}/{4}$ at other edge cells.


\subsubsection{{Magma movement}}
\label{sect:Piegari}

{\citet{Piegari2008,Piegari2011} developed the fracture model 
into a model for volcanism by adding a magma field $n_{i,j}$.} 
In the {original model \citep{Piegari2008}}, 
$n_{i,j}=0$ corresponds to no magma, 
while $n_{i,j}=1$ indicates a magma filled cell.
({A more refined model was introduced in \citet{Piegari2011};
this is discussed in section~\ref{sect:volatiles}.})
All magma is assumed to originate from an unlimited reservoir (fed from a deeper supply), which {is not modelled} explicitly. The top of this magma reservoir occupies the central quarter of the external row below {the} grid, $i=L+1$. Magma can enter the grid itself (the `magma chamber') when any cell next to the reservoir becomes fractured as a result of the evolving stress field. Once inside the magma chamber, magma migrates through the fracture network towards the surface (which corresponds to the external row above our grid, $i=0$). An eruption is said to occur once magma reaches a fractured cell adjacent to the surface (in row $i=1$). While large avalanches of fracture events do occur, most fractures are localised small-scale events, during which batches of magma can only move relatively small distances through the chamber. 
The motion of magma through the system will therefore generally be rather gradual, taking the form of a `diffusive' rise through the self-organised fracture network.

\par
\begin{figure*}[tbp]
   \centering
   \vspace*{-25mm}
   \includegraphics[scale=0.7]{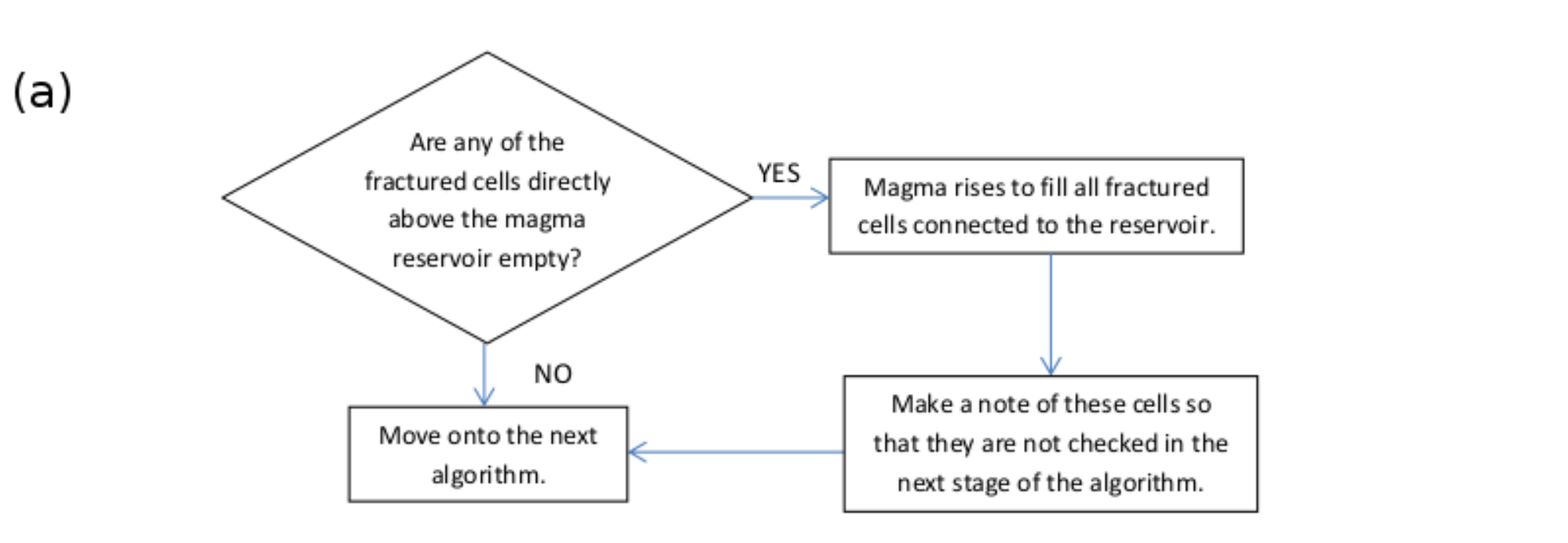}

   \vspace*{2.5mm}
   \includegraphics[scale=0.7]{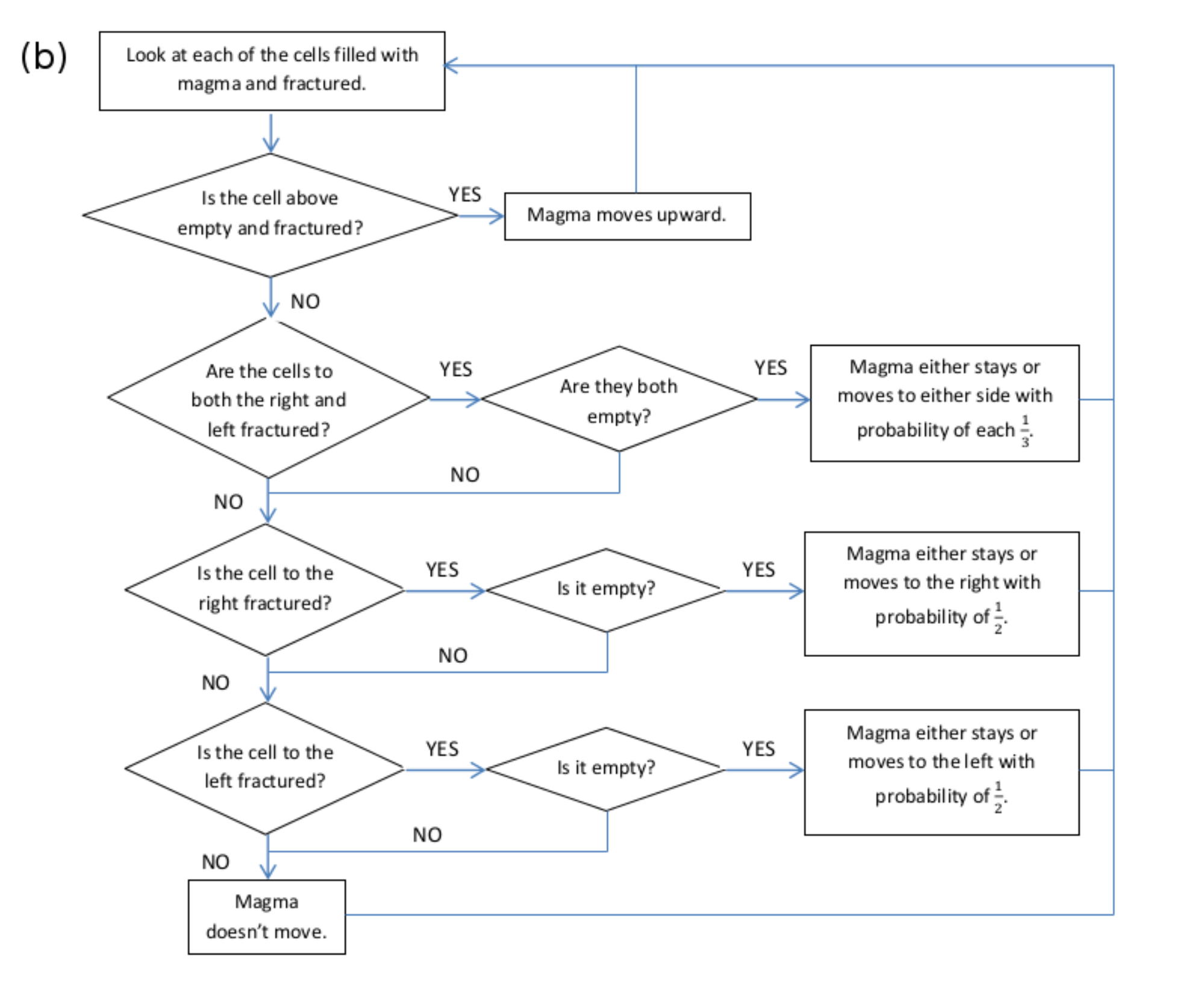}

   \vspace*{2.5mm}
   \includegraphics[scale=0.7]{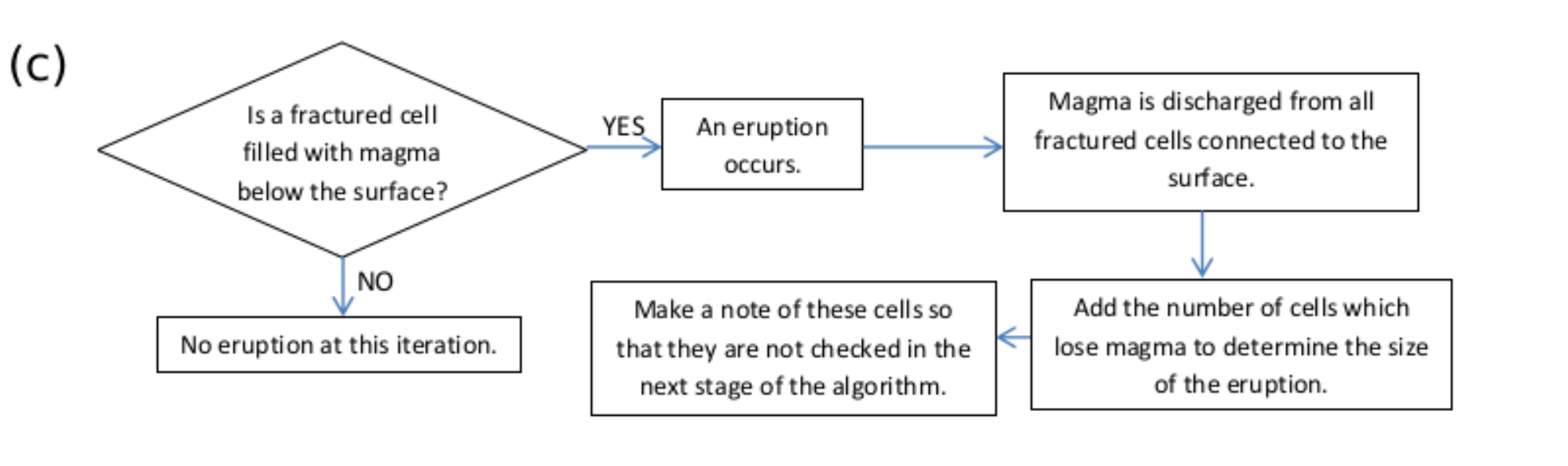}
   \caption{Flow diagrams describing: 
           (a) the rise of magma from the reservoir into the chamber;
           (b) the movement of magma within the chamber;
           (c) eruptions from the magma chamber to the surface.}
   \label{fig:flowcharts}
\end{figure*}

 The detailed algorithms controlling the magma movement are explained in the flow diagrams in Figure~\ref{fig:flowcharts}. After each fracture event, the movement of magma from the reservoir to the chamber is considered first (panel a). Magma is then allowed to move through the fracture network (panel b). In implementing the magma movement algorithm, we repeatedly iterate over all cells within the grid: looping vertically down columns ($i=1, \ldots L$) within  a horizontal loop across rows ($j=1, \ldots L$). The possibility of vertical movement due to buoyancy is considered throughout the grid before any horizontal movement is allowed. (Diagonal motion is not permitted.) 
We run this magma movement algorithm until no further vertical motion can take place, checking for eruptions 
(panel c) 
after each iteration of the algorithm. At this point, the current phase of magma movement is considered to be complete. The fractured cells close up, trapping any magma they contain, and the system evolves in time to the next fracture event (following the rules of section~\ref{sect:OFC}). 

\par Most eruptions only involve magma that has ascended through the magma chamber during multiple fracture events (and there are usually many fracture events between each eruption). However, a small number of larger, explosive eruptions can involve the reservoir directly; these would model events such as those at Soufri\`{e}re Hills in Monserrat, 
considered by \citet{Scandone2009}. In these unusual cases, where a fracture network
directly connects the reservoir to the surface,
the reservoir algorithm fills all such cells with magma, 
and {the} eruption algorithm immediately ejects this magma in an eruption.
To avoid an infinite loop,
only a single filling of the network from the reservoir takes place.
Physically, {the eruption is considered} to have caused the collapse
of the chamber walls surrounding the fracture network, 
preventing the rise of additional magma \citep[e.g.][]{Scandone2009}.
Note that such events would involve volatile-rich magma (see below) so will be particularly explosive.


\subsubsection{{Volatiles and Magma Degassing}}
\label{sect:volatiles}

In modelling volcanic explosivity, it is important to consider 
the volatile content of the magma,
which can greatly affect the style of eruption.
{\citet{Piegari2011} introduced this concern, assuming}
that water is the dominant volatile element
(although other volatiles might be treated similarly).
In equilibrium, the dissolved water concentration, $n_{\rm d}$, 
is determined by the lithostatic pressure $p$, as
\begin{equation}
\label{eq:Relationship}
n_{\rm d} = 6.8 \times 10^{-8} \; p^{0.7} \, ,
\end{equation}
with $n_{\rm d}$ the fractional water content, and $p$ the pressure in Pascal
\citep{Piegari2013},
with the lithostatic pressure calculated as
\begin{equation}
\label{eq:Relationship2}
p(z) = p_{0} + g \int_0^z \rho(\zeta)\,\mathrm{d}\zeta \, ,
\end{equation}
where $\rho(z)$ is the density of the rock at depth $z$,
and $g = 9.81\,{\rm m}{\rm s}^{-2}$ is the gravitational acceleration.
Thus the equilibrium value of $n_{\rm d}$ varies with depth, 
following the pressure variation.
To replicate the model of \citet{Piegari2011},
we adopt a grid of depth $h = 12\, \rm{km}$, 
surface pressure $p_{0} = 0$,
and constant rock density $\rho(z)=\rho_0=2700\, {\rm kg}{\rm m}^{-3}$. 

{The dissolved water content at the depth of the reservoir, 
$n_{0}$, is taken as} $n_0\approx0.06$, or 6\%. 
The gas lost due to exsolution, $n_{\rm loss}$, is calculated as
\begin{equation}
\label{nloss}
n_{\rm loss} = n_0 - n_{\rm d} \, .
\end{equation}
Thus moving from the reservoir to the surface 
(assuming local equilibration), 
$n_{\rm d}$ decreases from 0.06 to 0, 
and $n_{\rm loss}$ increases from 0 to 0.06.
We follow \citet{Piegari2011} in considering as our magma field the quantity
$n=(1 - n_{\rm loss})$, 
which varies from 1 at the reservoir to 0.94 at the surface.

Water is exsolved (magma is equilibrated)
if and only if the magma remains in the magma chamber 
between earthquake events, 
with $n_{i,j}$ then taking the value of $(1-n_{\rm loss})$ 
calculated for the depth at which the magma resides.
This corresponds to `passive degassing' \citep{NewhallToG},
with the volatives escaping slowly 
through the surrounding country rock,
and leaving the emplaced magma depleted in volatives
(and hence more viscous, less explosive), at the new equilibrium level.
When magma migrates through the fracture network during a fracture event,
it initially retains its volatile content.
So when an eruption occurs, 
the magma reaching the surface will have a range of volatilities, 
reflecting the depths at which the various batches of magma last resided. Rarely, this will involve some saturated magma, which has ascended directly from the reservoir during that earthquake event.
The mean volatility content of the magma taking part in an eruption is a proxy for the explosivity.


\subsection{{New model: magma-induced stress}}
\label{sect:delta}

{In the earlier model described above}, 
the stress field has been an independent quantity,
entirely determined by the OFC model, 
without any feedback from the magma field.
This models a uniform background rate of stress, without local variations.
But as discussed in section~\ref{sect:intro},
there are many reasons why magma intrusions should lead to 
enhanced local stress values.
Here we adapt the OFC model to impose additional stress 
on cells containing magma.

\par To model an enhanced rate of stress in those cells as simply as possible,
we increase the rate of stress in the magma-filled cells 
by a factor $(1+\delta)$,
where $\delta$ is a new model parameter.
We therefore increase the local strain rate from $\nu$ to $\nu(1+\delta)$.
Since we have normalised time with respect to $\nu$ 
(see section~\ref{sect:OFC}),
the stress in magma-filled cells will therefore increase by
$(1+\delta)\Delta t$ over a time interval $\Delta t$ 
(during which time the stress in other cells 
will simply increase by $\Delta t$). 
To proceed directly to the next fracture event, as before,
we need to consider the evolution of stress in cells with and without magma 
($f^{\rm mag}_{i,j}$ and $f^{\rm nomag}_{i,j}$, respectively)
separately.
{Note that $f^{\rm mag}_{i,j}$ and $f^{\rm nomag}_{i,j}$ 
are not parameterised in any way.
They are simply the original stress field $f_{i,j}$,
but now treated in one of two different ways, 
depending on whether or not cell $(i,j)$ contains magma.}

The time until the next fracture event in a cell without magma 
would be \begin{equation}
\label{eq:nomagstress}
\delta t_{\rm nomag} = 1 - {\rm max}(f^{\rm nomag}_{i,j}) \, ;
\end{equation}
the time until the next fracture event in a cell containing magma
would be
\begin{equation}
\label{eq:magstress}
\delta t_{\rm mag} = \frac{1 - {\rm max}(f^{\rm mag}_{i,j})}{1 + \delta} \; .
\end{equation}
The next event therefore occurs after the lesser of these two times,
\begin{equation}
\delta t = {\rm min}\left(\delta t_{\rm nomag}, \delta t_{\rm mag} \right)\; ,
\end{equation}
and we can then update the stresses as
\begin{equation}
f^{\rm nomag}_{i,j} \longrightarrow f^{\rm nomag}_{i,j} + \delta t  \; , \qquad
f^{\rm mag}_{i,j} \longrightarrow f^{\rm mag}_{i,j} + (1 + \delta) \, \delta t  \; .
\end{equation}

\par {Note that the addition of magma-induced stress only directly affects
the fracture model (the OFC model), 
which is now coupled to the magma model as described above;
the algorithms for magma movement and degassing remain unchanged.
Nevertheless, the resulting eruption behaviour can differ markedly,
in the presence of the different styles of fracture networks 
obtained in the coupled system.
The fracture dynamics (and therefore also the eruption dynamics)
of course depend on the value of our new parameter $\delta$;
the outcome for varying values of $\delta$ is discussed 
in section~\ref{sect:results}.}


\subsection{Numerical Implementation}

The model described above was implemented computationally in Fortran.
Although some larger grid sizes were considered, we performed most runs
on a grid of size $L=64$.  
With the grid representing a length of 12 km in both directions, 
we therefore consider a cell spacing of 200 m.
We typically perform runs involving $10^{8}$ eruptions
and analyse the statistics of the ensuing volcanic activity
with respect to distributions of event size, repose time, 
and explosivity.
{For reasons of computational efficiency, 
the data for $10^8$ eruptions are actually compiled from 40 individual runs, 
each of $2.5 \times 10^{6}$ eruptions.
For each individual run, 
all data associated with the first $10^{5}$ eruptions were discarded,
with the consequence that all data analysed are from systems 
that have attained statistically steady states.}


\section{Results}
\label{sect:results}


\subsection{{Earlier model: Piegari et al.\ (2008, 2011)}}
\label{sect:original}

Before investigating 
{the detailed effects of the new magma-induced stress mechanism
described in section~\ref{sect:delta}},
we reproduce the behaviour described by \citet{Piegari2008, Piegari2011}
for their model (corresponding to our case $\delta=0$).
We had previously verified our implementation of the underlying OFC model with the results
of \citet{Olami1992}.

\begin{figure}[tbp]
        \centering
        \includegraphics[scale=0.5]{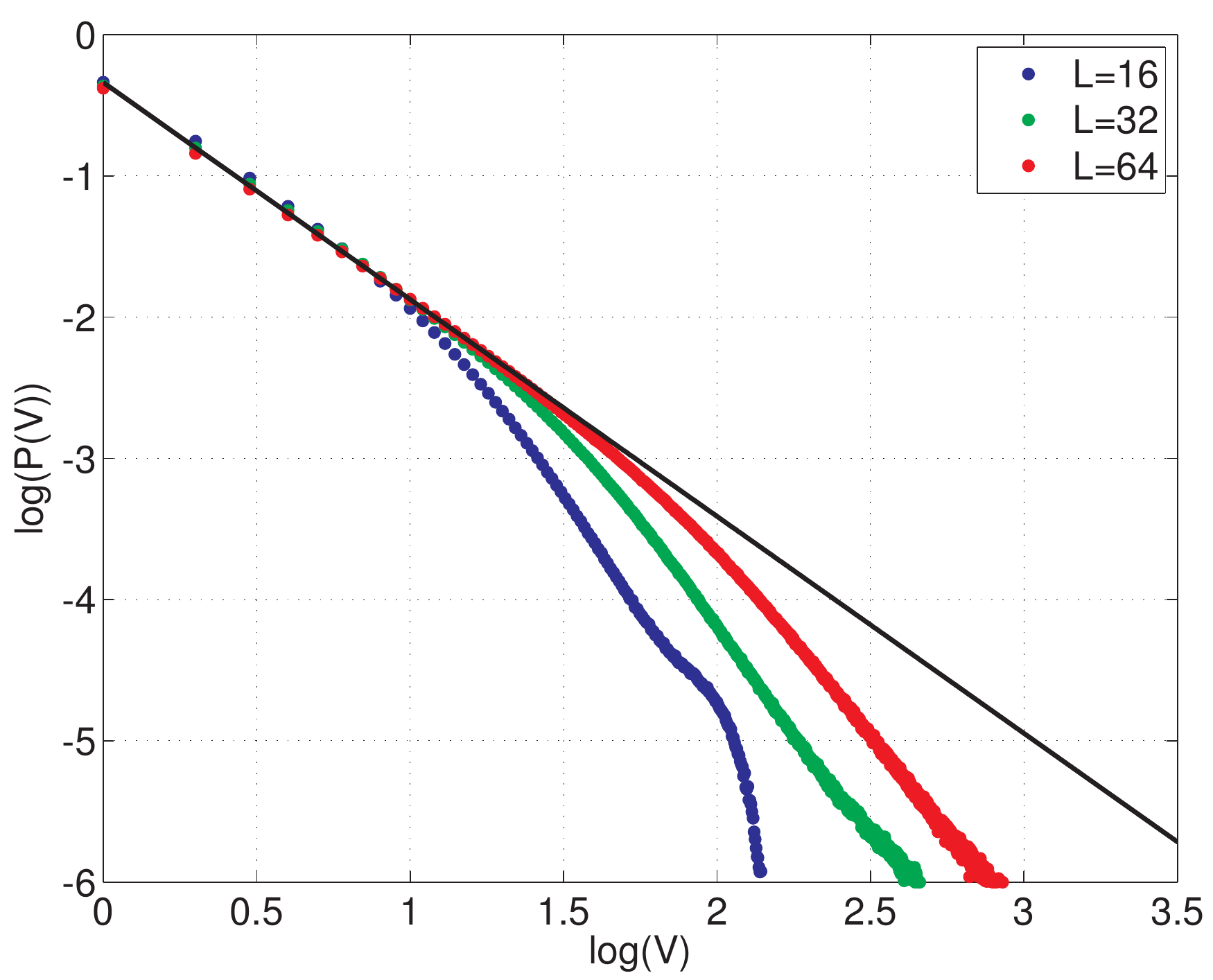}   
        \caption{Probability distribution $P(V)$ of eruptions of size $V$,
         for varying $L$ from runs with $10^8$ eruptions.
         The straight line corresponds to the power law fit 
         $P(V)=V^{-\alpha}$ with $\alpha\simeq 1.49$.}
        \label{fig:PV}
\end{figure}

\par Figure~\ref{fig:PV} shows the probability density $P(V)$
of eruptions of size $V$ (i.e. the number of cells with magma involved in the eruption), on a log-log scale, 
from runs involving $10^8$ eruptions in systems of varying size $L$.
(We use $\log_{10}$ throughout.)
The straight line fit reflects a power law behaviour,
$P(V) \propto V^{-\alpha}$,
mirroring the observed fit to VEI values of \citet{Simkin1993}.
The power law behaviour breaks down at larger event sizes,
with the point of divergence being somewhat dependent upon
the system size (because the larger events are not
possible within smaller systems). Figure~\ref{fig:PV} should be compared
with Figure~2 of \citet{Piegari2008}.
They quote a value $\alpha=1.6$ for their power law exponent, but inspection of their figure suggests a value closer to $1.4$. Our straight line fit,
calculated for the range $0 \le \log V \le 1$,
corresponds to an exponent of $1.49$. 
Given the uncertainties in the fit
(the fit varies slightly with the range over which it is calculated,
as well as having a formal uncertainty)
and the possibility of minor differences between the fine details of our magma movement algorithms, we consider this agreement acceptable.

\par
\begin{figure}[tbp]
       \centering
        \includegraphics[scale=0.5]{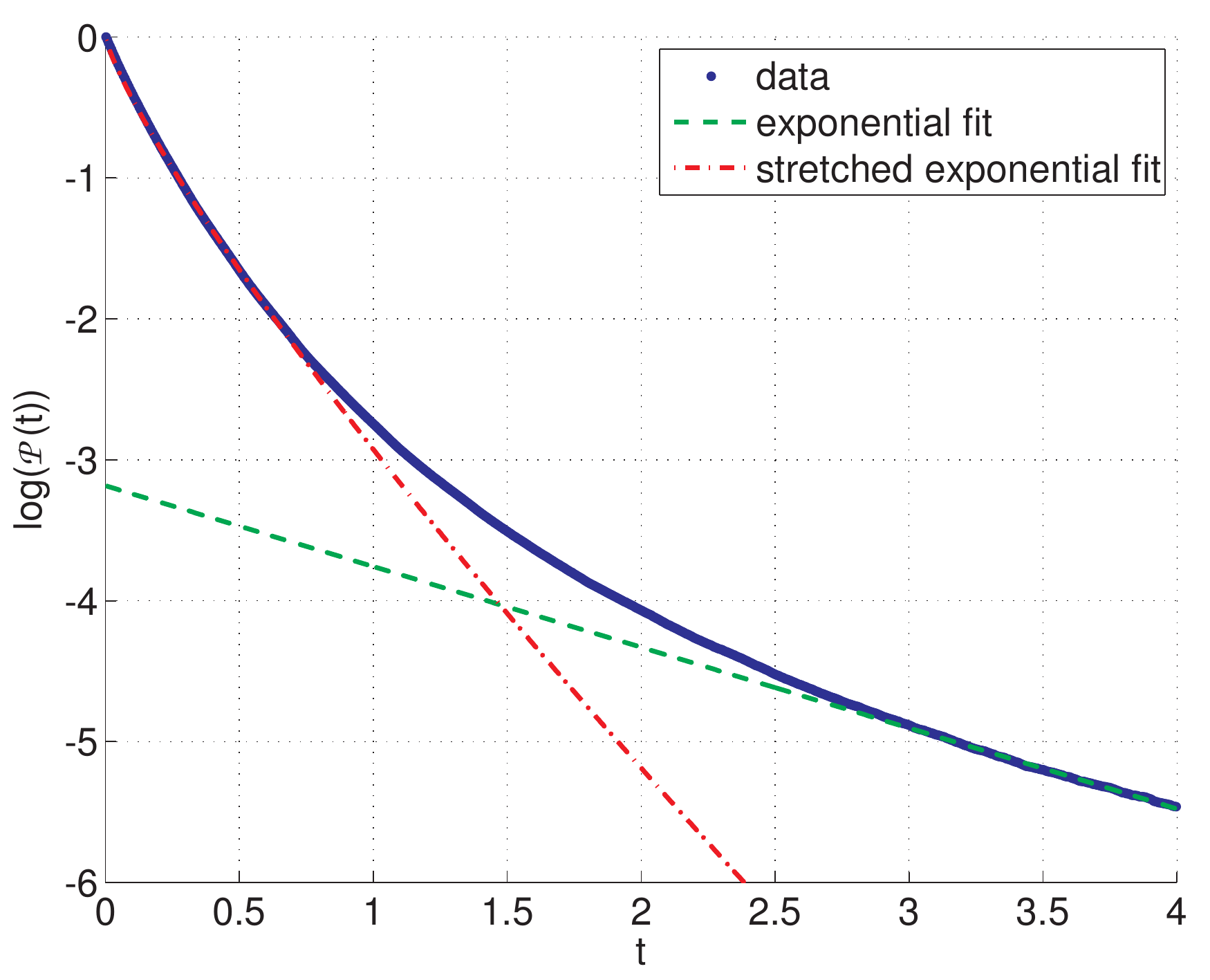}
        \caption{Cumulative probability distribution ${\cal P}(t)$ 
         of inter-eruption times (blue dots), 
         calculated for $10^8$ events with $L=64$.  
         An exponential fit to data in the last quarter of this range 
         (red line), and a stretched exponential fit to data in the first 
         quarter of this range (magenta line) are also shown.}
        \label{fig:Pt}
\end{figure}

 Figure~\ref{fig:Pt} is a log-linear plot of the cumulative probability ${\cal P}(t)$,
more formally ${\cal P}(T\ge t)$,
of inter-eruption intervals (repose times) of duration $T$, 
for the same model as above. Note that this is the complement of the more conventional 
cumulative probability $\overline{\cal P}(T\le t)$, 
related by ${\cal P}(T\ge t)=1-\overline{\cal P}(T\le t)$;
thus ${\cal P}(0)=1$ and ${\cal P}(t)\longrightarrow0$ 
as $t\longrightarrow\infty$. This can be compared to Figure~3 of \citet{Piegari2008}.
In both cases the time units are non-dimensional
(see section~\ref{sect:Piegari}),
and the plots exclude the longest repose times, 
in the tail of the distribution.

\par For large repose times (i.e.\ for 3 $\le t \le$ 4) the ${\cal P}(t)$ curve in Figure~\ref{fig:Pt} approaches a linear slope, corresponding to exponential behaviour, ${\cal P}(t) \propto \exp(-t/\tau_1)$. Here  $\tau_1 \simeq 0.76 \pm 0.06$, with the uncertainty in the fit corresponding to variations due to different choices of range. Exponential behaviour of this type is the expected result for a Poisson process, indicating that such events are `memoryless' (i.e.\ independent of previous similar events). This exponential behaviour only occurs for the long repose-time events towards the tail of the distribution (accounting for less than 0.1\% of the total events). The majority of eruptive events have a much shorter repose time. For such smaller repose-time events, the data is better approximated by a stretched exponential fit of the form ${\cal P}(t) \propto \exp\left(-(t/\tau_0)^\beta\right)$, where $\beta \simeq 0.83$ and $\tau_0 \simeq 0.10$. 
(The smaller events are therefore not well modelled as a Poisson process.
This is reasonable for the model of volcanism {considered} here,
with magma moving through the systems in small batches,
and consecutive events being linked via the dynamics of magma transport;
we do not expect memoryless behaviour for such events.)
{These results are qualitatively similar to those} reported by \citet{Piegari2008}.
However, there is some uncertainty in how their timescale relates to ours: \citet{Piegari2008} report their timescale in units of $\nu^{-1}$ (and state that their results depend on $\nu$), whereas we have scaled $\nu$ out of our calculations. A factor of $10^4$ difference in the respective timescales would account for differences between the reported fitting parameters. In any event, our agreement with Piegari et al.\ is qualitatively satisfactory, even if the quantitative comparison remains unclear.

\par
\begin{figure}[tbp]
        \centering
        \includegraphics[scale=0.48]{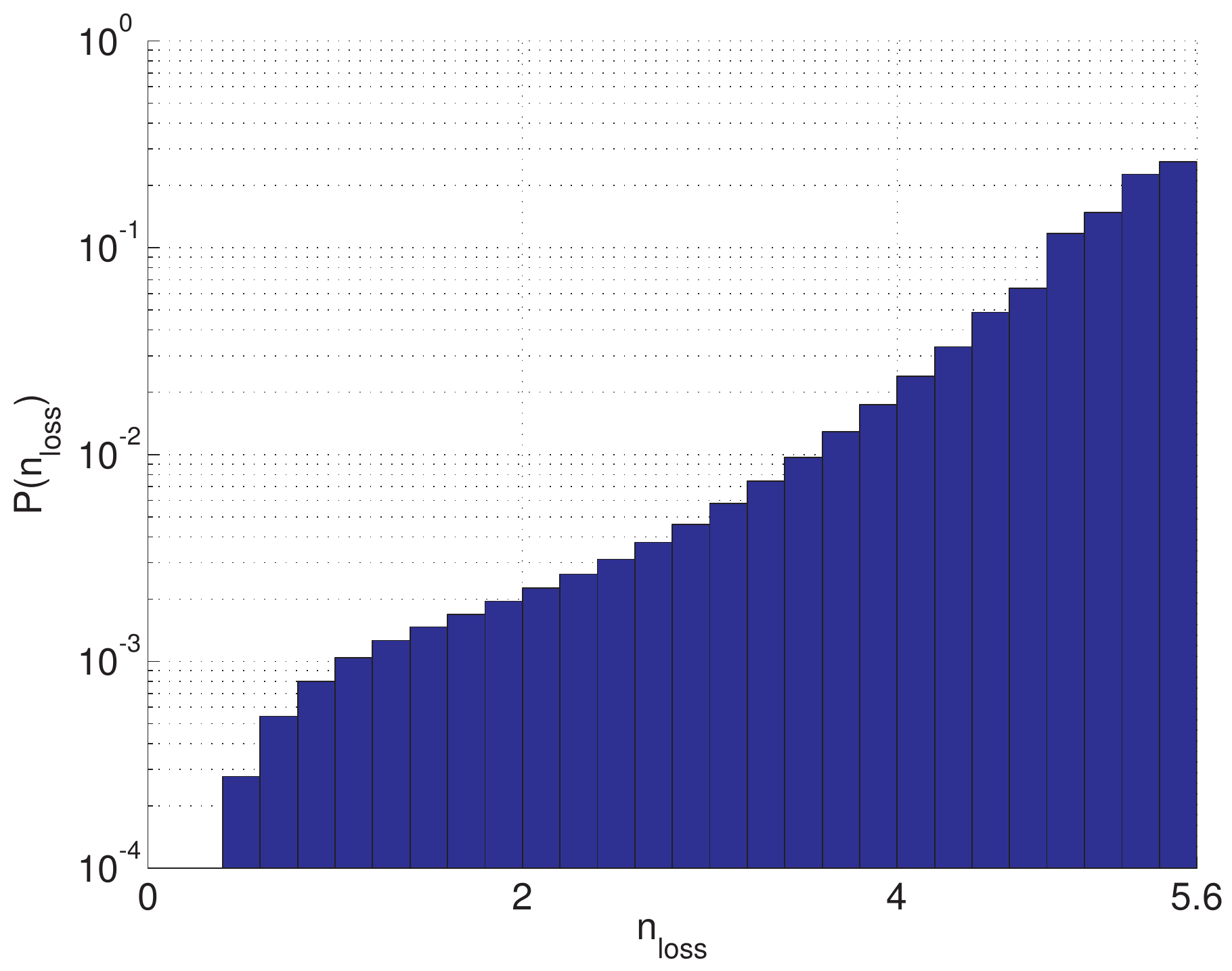}
        \caption{Probability distribution $P(n_{\rm loss})$ 
         of eruptions with percentage of gas loss $n_{\rm loss}$,
         calculated for $10^8$ events with $L=64$.}
	\label{fig:Pnloss}
\end{figure}

 The behaviour of volatile components within {the} model (and the associated range of explosivities of the eruptions) can also be compared with the calculations of \citet{Piegari2011}. Figure~\ref{fig:Pnloss} shows a histogram of the probability density $P(n_{\rm loss})$ of events with mean gas-loss $n_{\rm loss}$, on a log-linear scale.
The eruption statistics were obtained using a bin width of 0.2\% in $n_{\rm loss}$. The histogram, which is in broad agreement with Figure~4 of \citet{Piegari2011}, shows the far greater probability 
of eruptions involving magma that has lost most of its gas content
(having degassed relatively close to the surface).
At the other end of the spectrum, 
eruptions involving almost fully saturated magma 
are extremely rare. Such eruptions, with significant amounts of magma erupting directly from the reservoir (or at least from locations near the base of the
magma chamber) are typically larger in size,
and more explosive. The distribution is approximately exponential 
for most events, showing a (broadly) linear trend on the log-linear plot.

\begin{figure*}[tbp]
        \centering

\begin{subfigure}{.45\textwidth}
\centering
\includegraphics[scale=0.325]{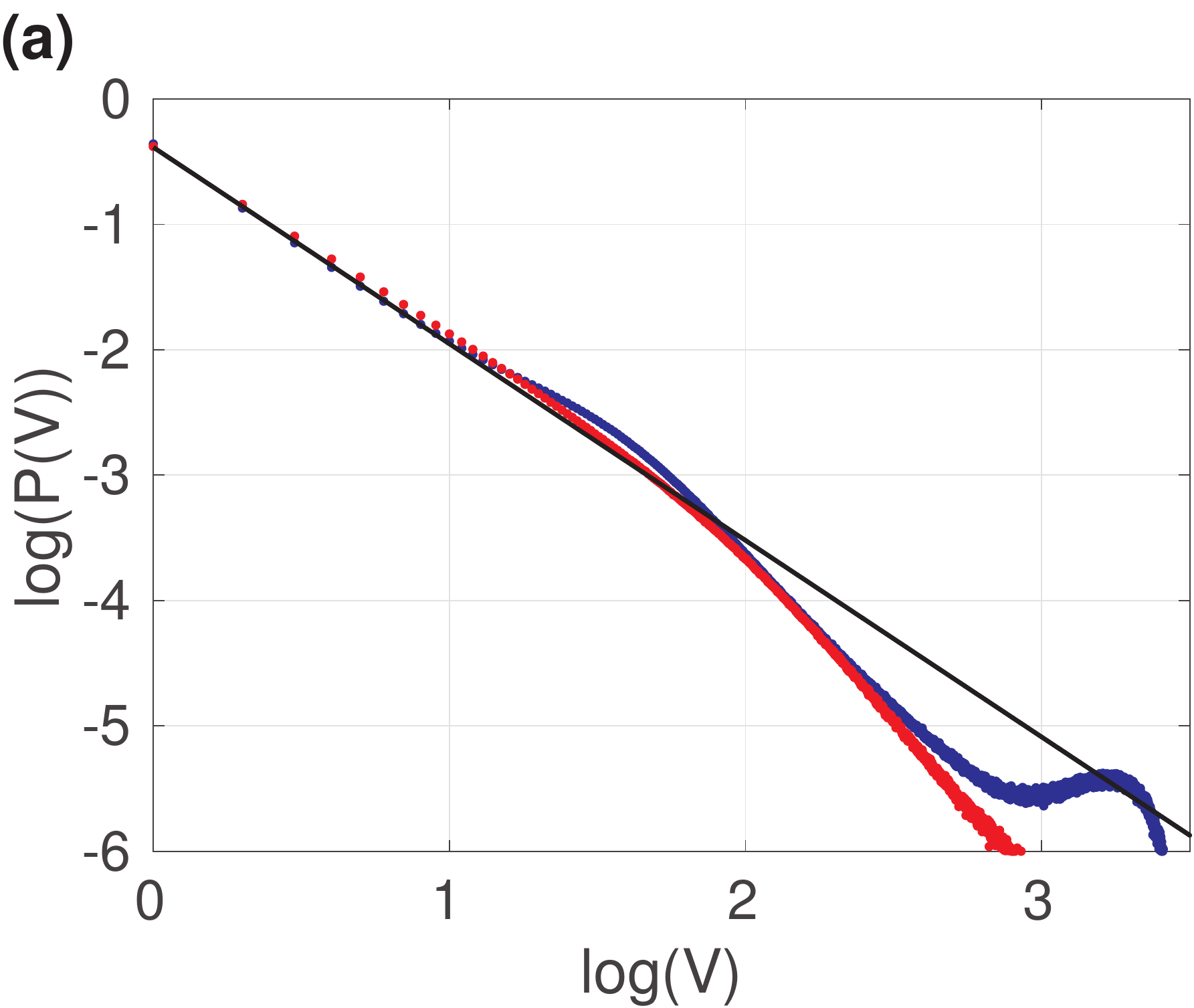}
\end{subfigure}%
\begin{subfigure}{.45\textwidth}
\centering
\includegraphics[scale=0.325]{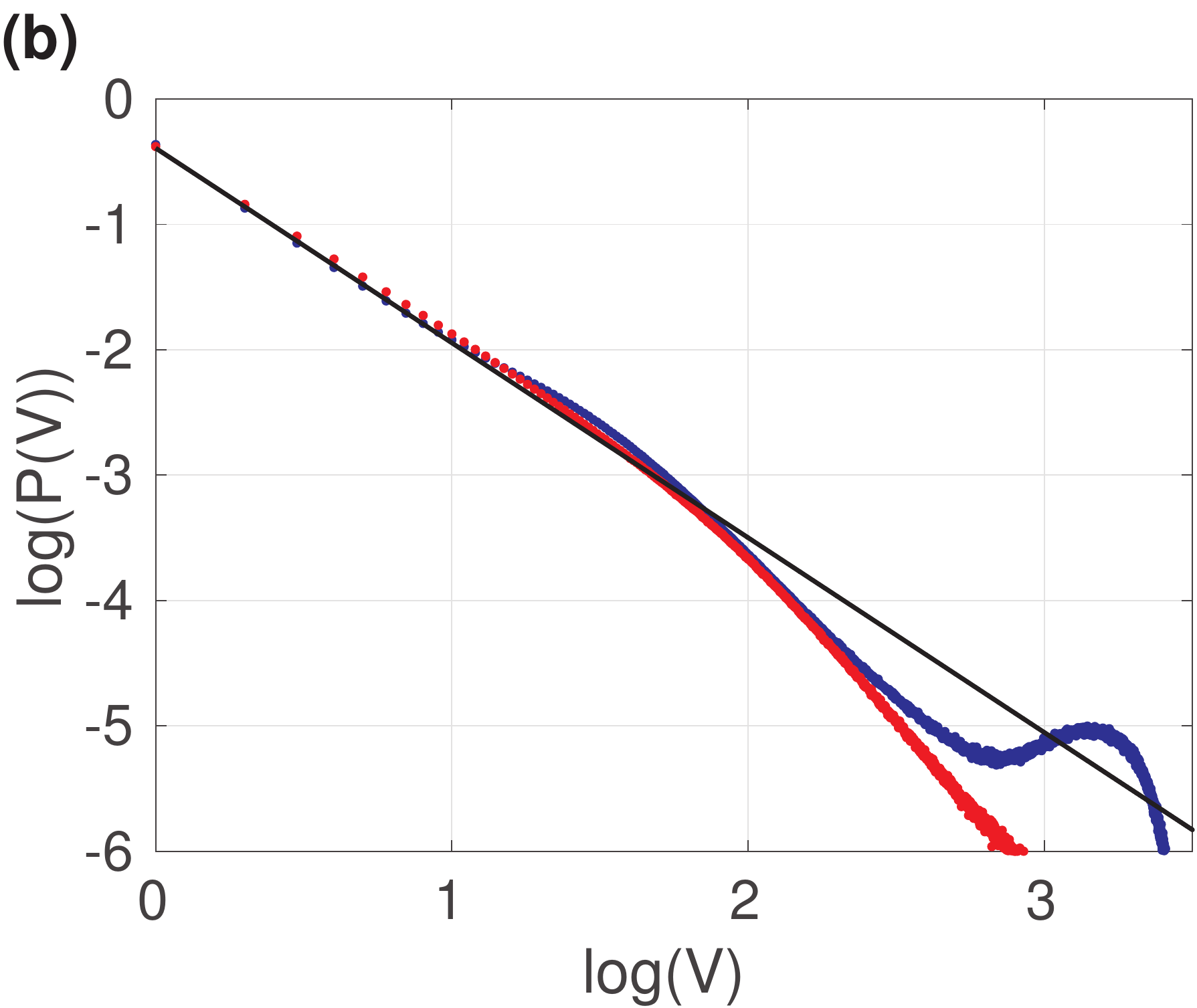}
\end{subfigure}%

\begin{subfigure}{.45\textwidth}
\centering
\includegraphics[scale=0.325]{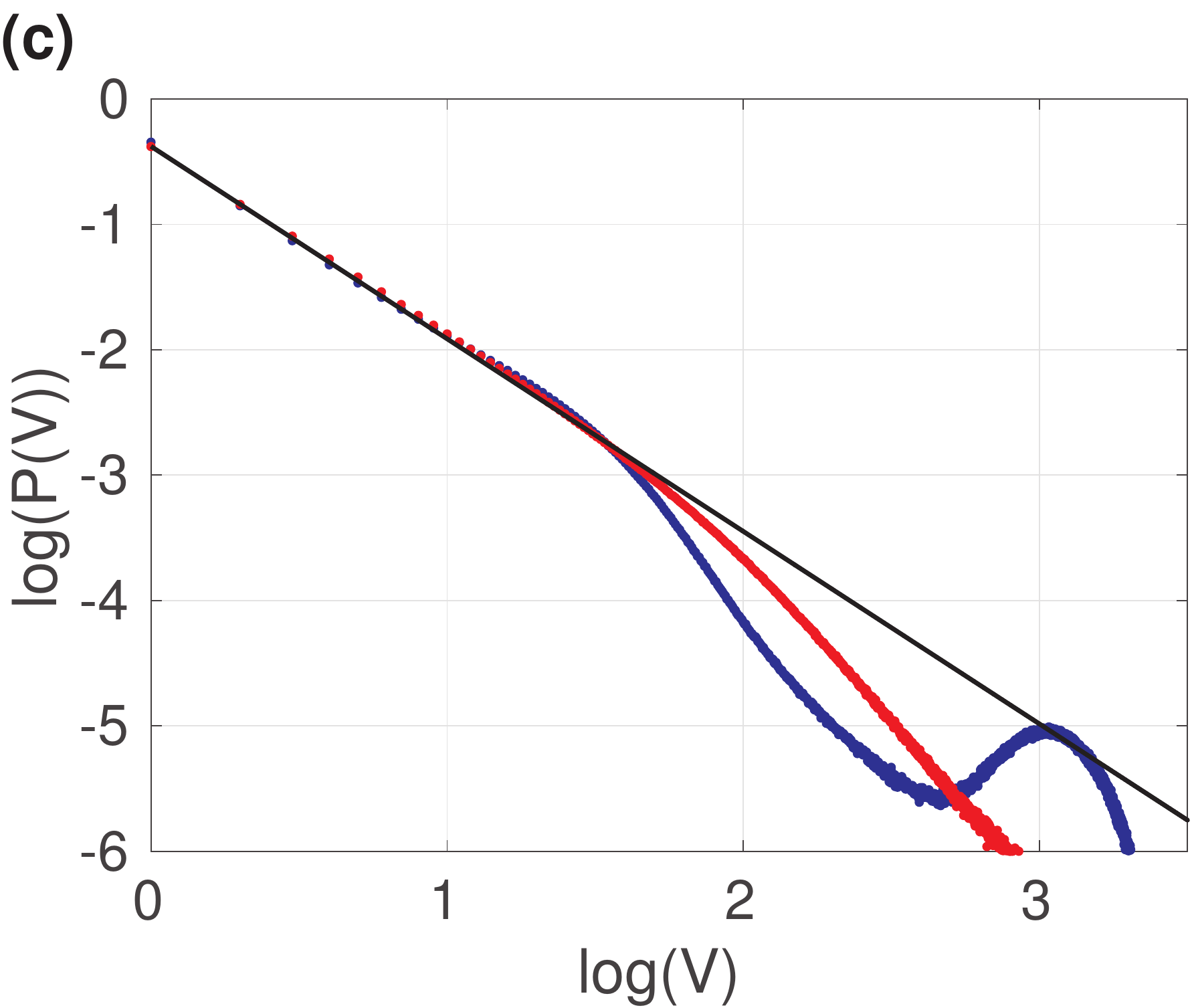}
\end{subfigure}%
\begin{subfigure}{.45\textwidth}
\centering
\includegraphics[scale=0.325]{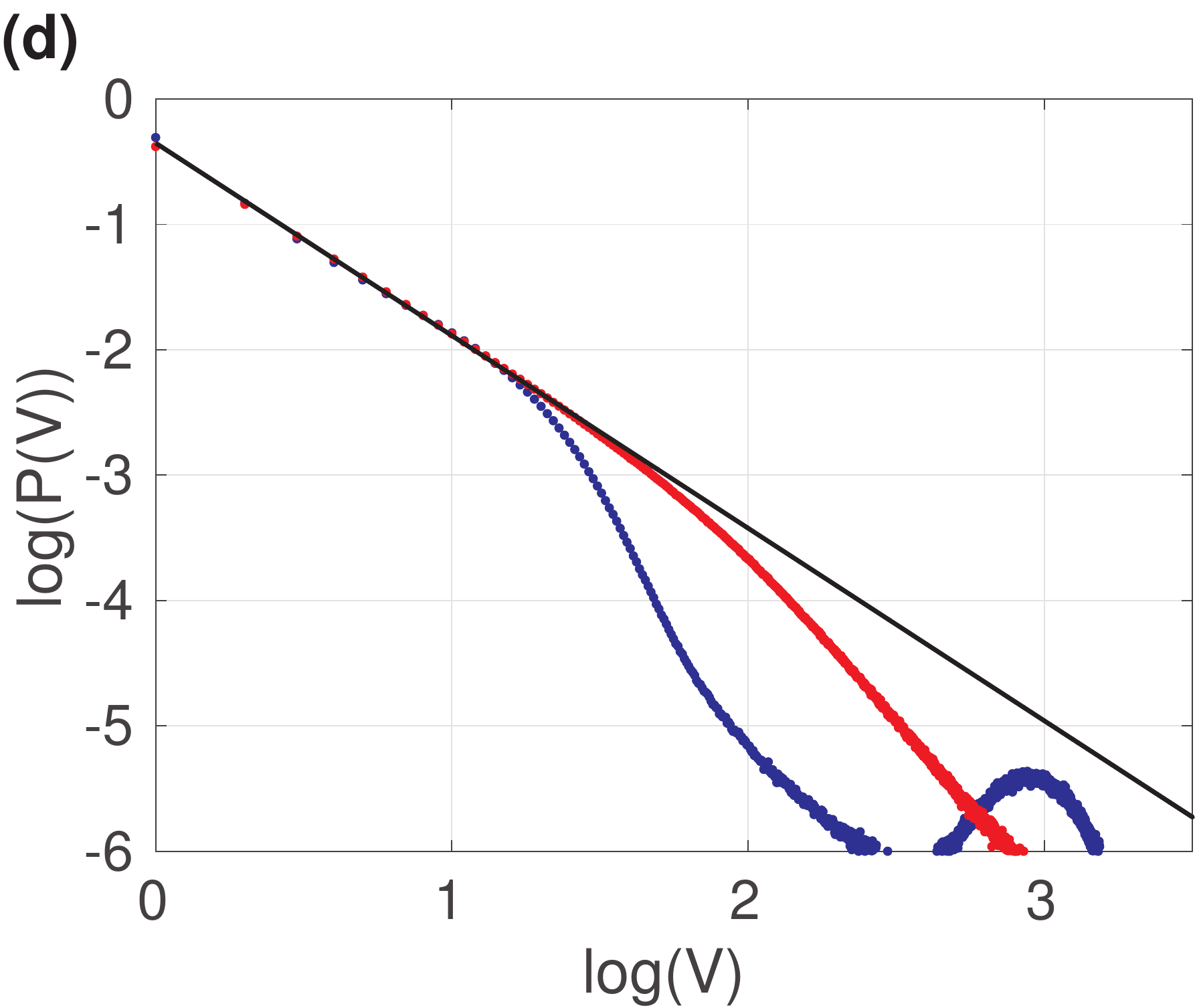}
\end{subfigure}%

        \caption{Probability distribution $P(V)$ for eruptions of size $V$, 
         using data from $10^8$ eruptions with $L=64$ for varying $\delta$ (in blue):
(a) $\delta=0.1$; (b) $\delta=0.2$; (c) $\delta=0.4$; (d) $\delta=0.6$.
         The straight lines are the power law fits 
         described in Table~\ref{tab:stats}.
         {The red data show the corresponding distribution for $\delta=0$,
         as shown in Figure~\ref{fig:PV}.}}
        \label{fig:PVdelta}
\end{figure*}

\subsection{{New model: magma-induced stress}}
\label{sect:magma}

We now consider the new effect of magma-induced stress ($\delta > 0$).
Figure~\ref{fig:PVdelta} shows the probability density functions $P(V)$
for various $\delta$ in the range 0.1--0.6,
from simulations with $L=64$ involving $10^8$ eruptions.
These can be compared with the corresponding distribution for $\delta=0$,
{also shown on each panel}.
%
The new functions $P(V)$ continue to show a
power law distribution for smaller events,
but the range over which this applies 
depends on $\delta$;
it decreases from $\log V\simeq 1.5$ with $\delta=0$,
to $\log V\simeq 1.1$ with $\delta=0.6$.
For intermediate $\delta$ ($0.1\le \delta \le 0.4$),
there is a local peak of events of moderate size
($\log V \simeq 1.5$) and, after a relative dearth of intermediate size events,
there is another prominent peak of large events
at $\log V\simeq 3.0\mbox{--}3.25$.
Note that for intermediate values of $\delta$,
both these peaks lie above the continuation of the power law line;
so, although obviously rarer than small events,
these eruptions must be considered relatively frequent.
It is possible that the extent of the peak at large $V$ 
is somewhat limited by the relatively small $L$ employed here,
and future calculations will investigate this.

\par The variation with $\delta$ of the mean eruption size $\overline{V}$,
and the slope and intercept of the power law fits, 
are given in Table~\ref{tab:stats}.
The behaviour with increasing $\delta$ is clear. As $\delta$ increases to 0.2, 
the mean eruption size $\overline{V}$ increases significantly.
Note that this coincides with a steepening of the power law slope, 
so this effect is clearly due to the localised peaks in the $P(V)$ distribution.
As $\delta$ increases further, these trends reverse;
the intermediate and large event size peaks become less significant,
and the main effect of the magma-induced stress
now seems to be an increased dominance of smaller events.
{These effects are due to the relative ease with which the system
allows magma to migrate between the reservoir and the surface,
with increasing $\delta$;
this is discussed further below, 
after additional aspects of the behaviour with varying $\delta$ 
have been discussed.}

\begin{table*}[tb]
\centering
\begin{tabular}{|l|c|cccccc|} \hline
$\delta$       & 0.0    & 0.1    & 0.2    & 0.3    & 0.4    & 0.5    & 0.6  \\ \hline
$\overline{V}$ & 9.80   & 19.5   & 28.5   & 22.8   & 16.3   & 11.2   & 7.01 \\
$\alpha$       & 1.49   & 1.57   & 1.55   & 1.54   & 1.54   & 1.53   & 1.54 \\
$P(V=1)$       & -0.384 & -0.385 & -0.392 & -0.385 & -0.378 & -0.367 & -0.351 \\ \hline
$\overline{T}$ & 0.115  & 0.157 & 0.140  & 0.124  & 0.119  & 0.120  & 0.127 \\
$\tau_0$       & 0.10   & 0.13  & 0.12  & 0.10  & 0.10  & 0.10  & 0.12 \\
$\beta$        & 0.83   & 0.79  & 0.79  & 0.80  & 0.82  & 0.85  & 0.89 \\ 
$\tau_1$       & 0.76   & 0.84  &  0.82 & 0.77  & 0.67  & 0.53  & 0.46 \\  \hline
\end{tabular}
\caption{Some eruptions statistics with varying $\delta$.
Mean eruption size $\overline{V}$, 
and the slope (-$\alpha$) and intercept ($P(V=1)$) 
from the power law fits in Figures~\ref{fig:PV}, \ref{fig:PVdelta}.
Mean inter-eruption time $\overline{T}$, 
and stretched exponential ($\tau_0$, $\beta$)
and exponential ($\tau_1$) parameters
from the fits in Figures~\ref{fig:Pt}, \ref{fig:Ptdelta}.}
\label{tab:stats}
\end{table*}

\par
\begin{figure*}[tbp]
        \centering

\begin{subfigure}{.45\textwidth}
\centering
\includegraphics[scale=0.325]{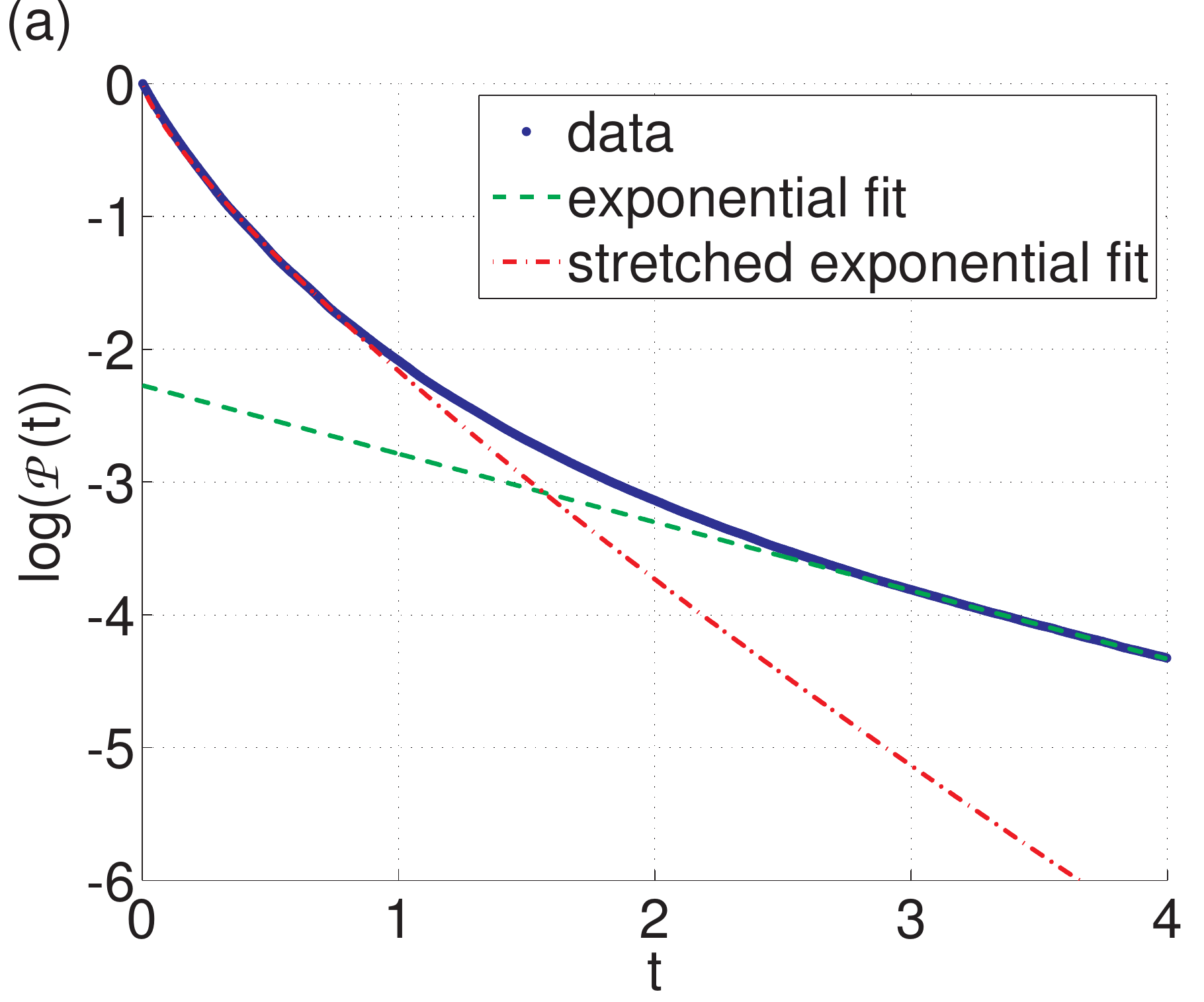}
\end{subfigure}%
\begin{subfigure}{.45\textwidth}
\centering
\includegraphics[scale=0.325]{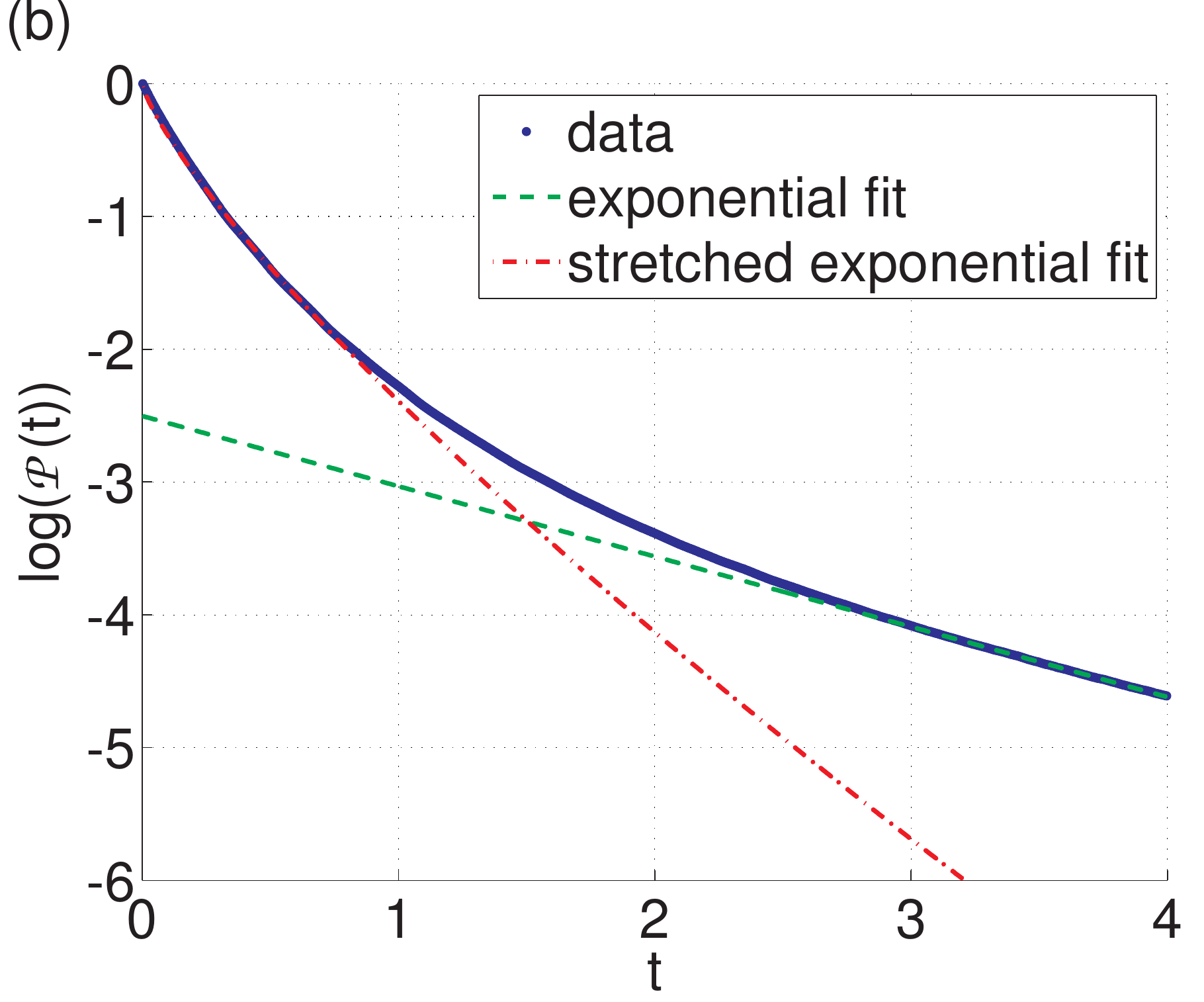}
\end{subfigure}%

\begin{subfigure}{.45\textwidth}
\centering
\includegraphics[scale=0.325]{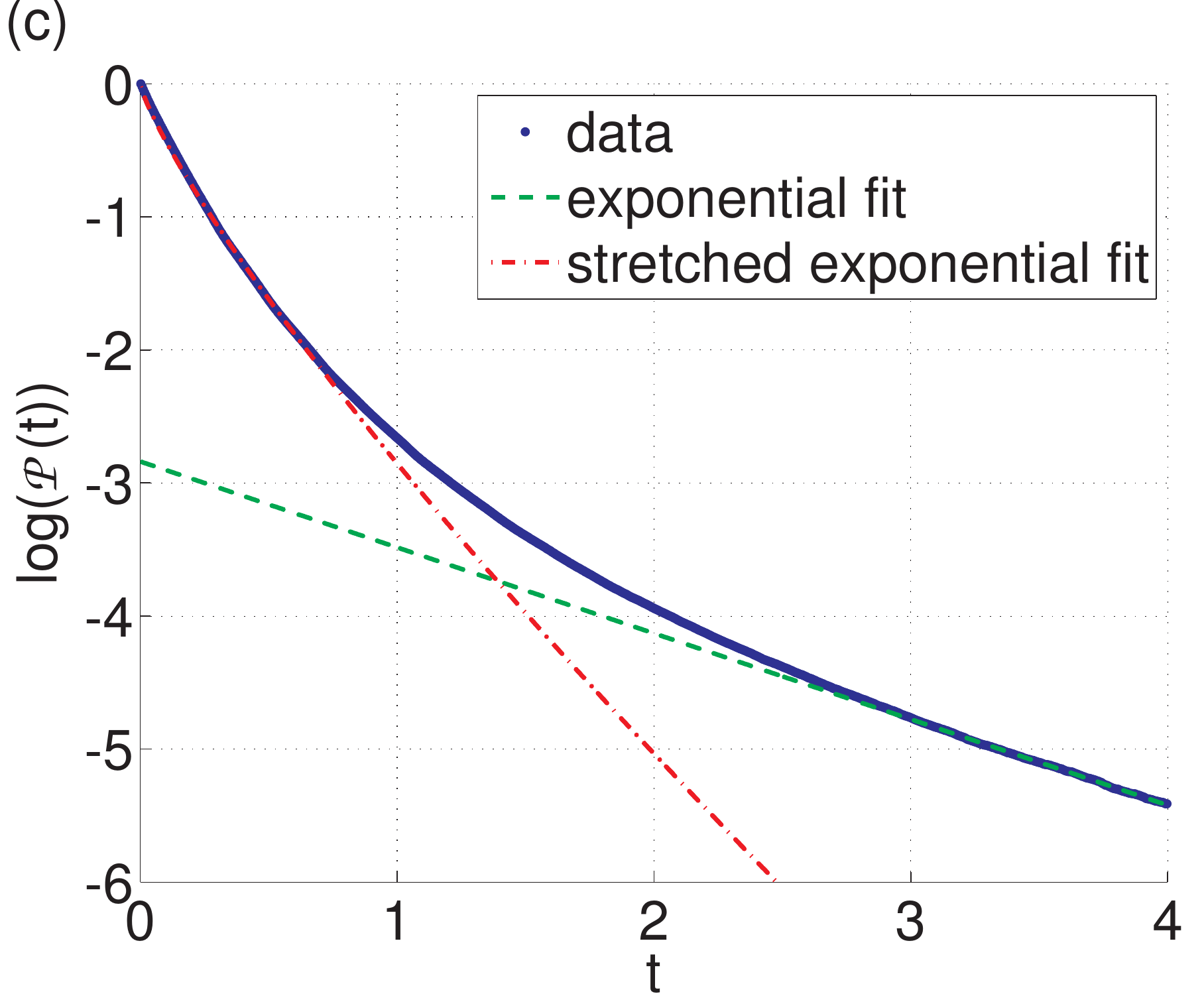}
\end{subfigure}%
\begin{subfigure}{.45\textwidth}
\centering
\includegraphics[scale=0.325]{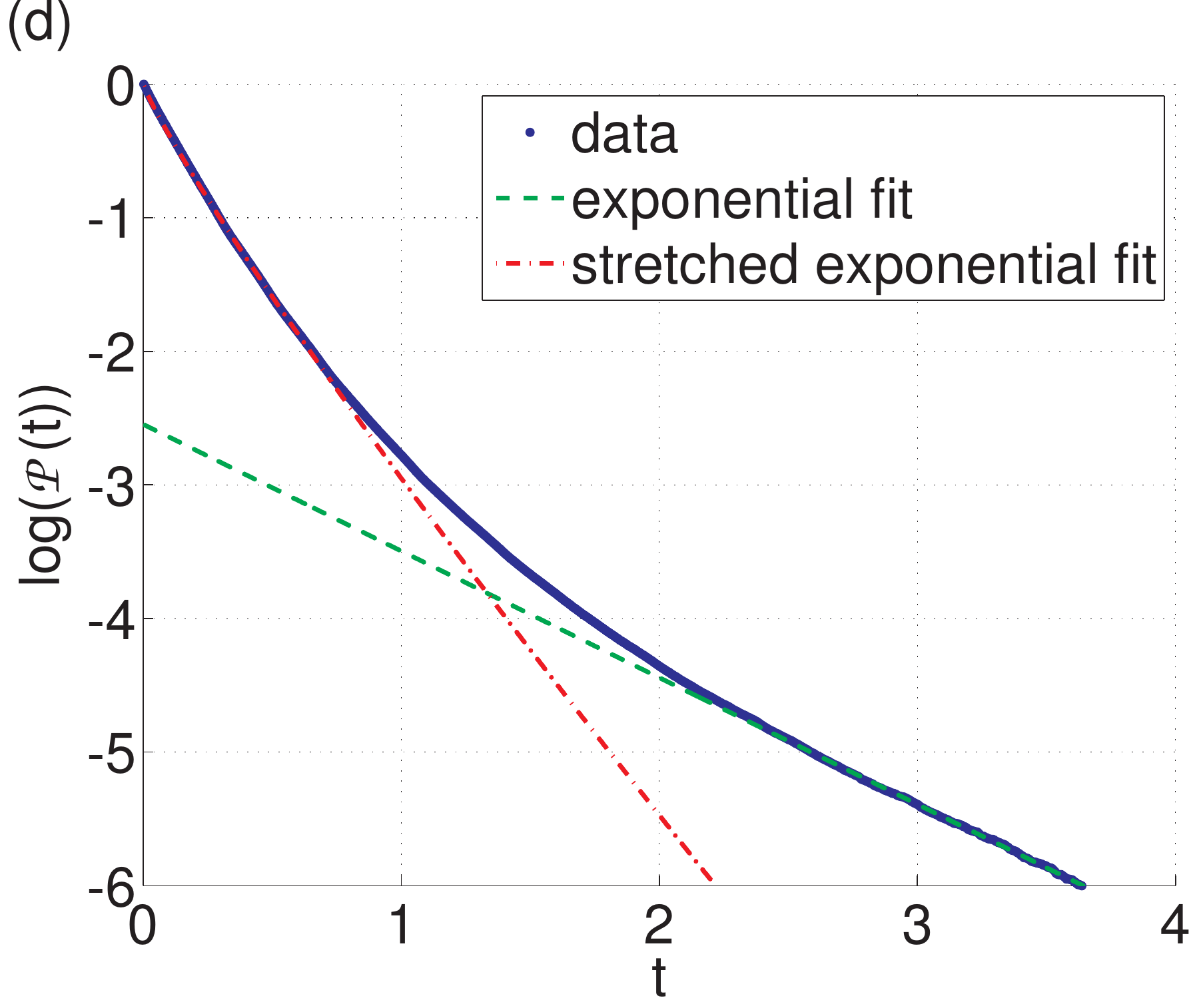}
\end{subfigure}%

        \caption{Cumulative probability distribution ${\cal P}(t)$ 
         of inter-eruption times,
         using data from $10^8$ eruptions with $L=64$ for varying $\delta$:
         (a) $\delta=0.1$; (b) $\delta=0.2$; (c) $\delta=0.4$; (d) $\delta=0.6$.
         The exponential and stretched exponential fits
         are described in Table~\ref{tab:stats}.}
        \label{fig:Ptdelta}
\end{figure*}

 The effects on the cumulative probability distribution 
of events ${\cal P}(t)$, are consistent with 
{the picture described above};
Figure~\ref{fig:Ptdelta} shows this distribution
for the same values of $\delta$ considered above.
(The corresponding distribution for $\delta=0$ 
is shown in Figure~\ref{fig:Pt}.)
The variation with $\delta$ of the mean inter-eruption time 
and the parameters of the stretched exponential and exponential fits
are also given in Table~\ref{tab:stats}.
As $\delta$ increases to 0.1, 
the mean inter-eruption time $\overline{T}$ increases,
due to the increasing prominence of intermediate and large eruptions
(which empty the central region of the magma chamber,
leading to significant times before subsequent eruptions).
The timescale $\tau_1$ for long inter-eruption times (from the exponential fit)
increases from 0.76 to 0.84.
For small inter-eruption times, $\tau_0$ also increases (from 0.10 to 0.13)
and the stretching exponent $\beta$ decreases slightly (from 0.83 to 0.79).
As $\delta$ increases further however, these trends again reverse ---
for $\delta\simeq 0.3$--0.4, the $P(t)$ distribution 
is broadly similar to that for $\delta=0.0$ ---
and by $\delta=0.6$, the shorter inter-eruption times
(predominantly associated with smaller eruptions)
increasingly dominate the $P(t)$ distribution,
with very long inter-eruption times increasingly rare.

\par
\begin{figure*}[tbp]
        \centering

\begin{subfigure}{.45\textwidth}
\centering
\includegraphics[scale=0.325]{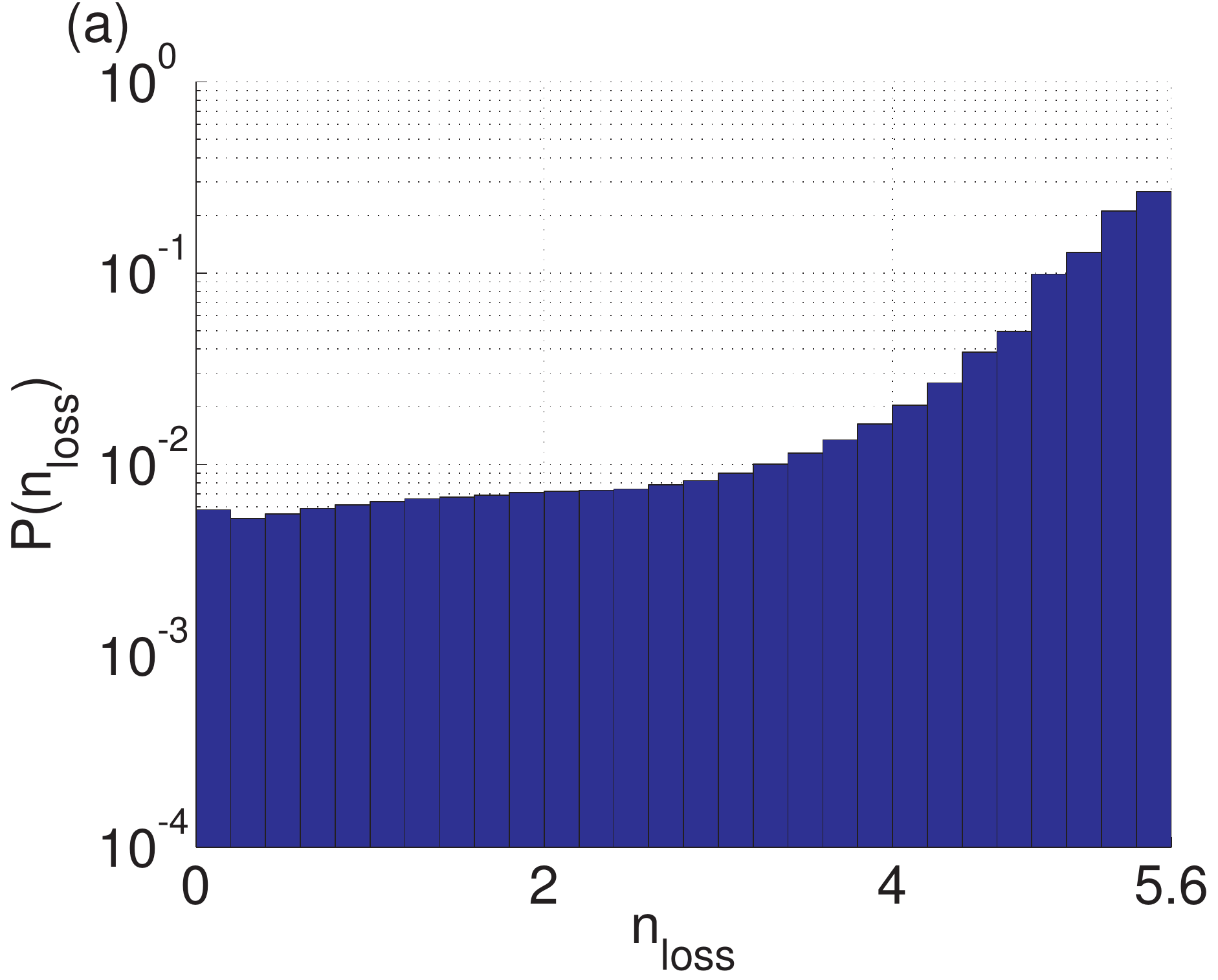}
\end{subfigure}%
\begin{subfigure}{.45\textwidth}
\centering
\includegraphics[scale=0.325]{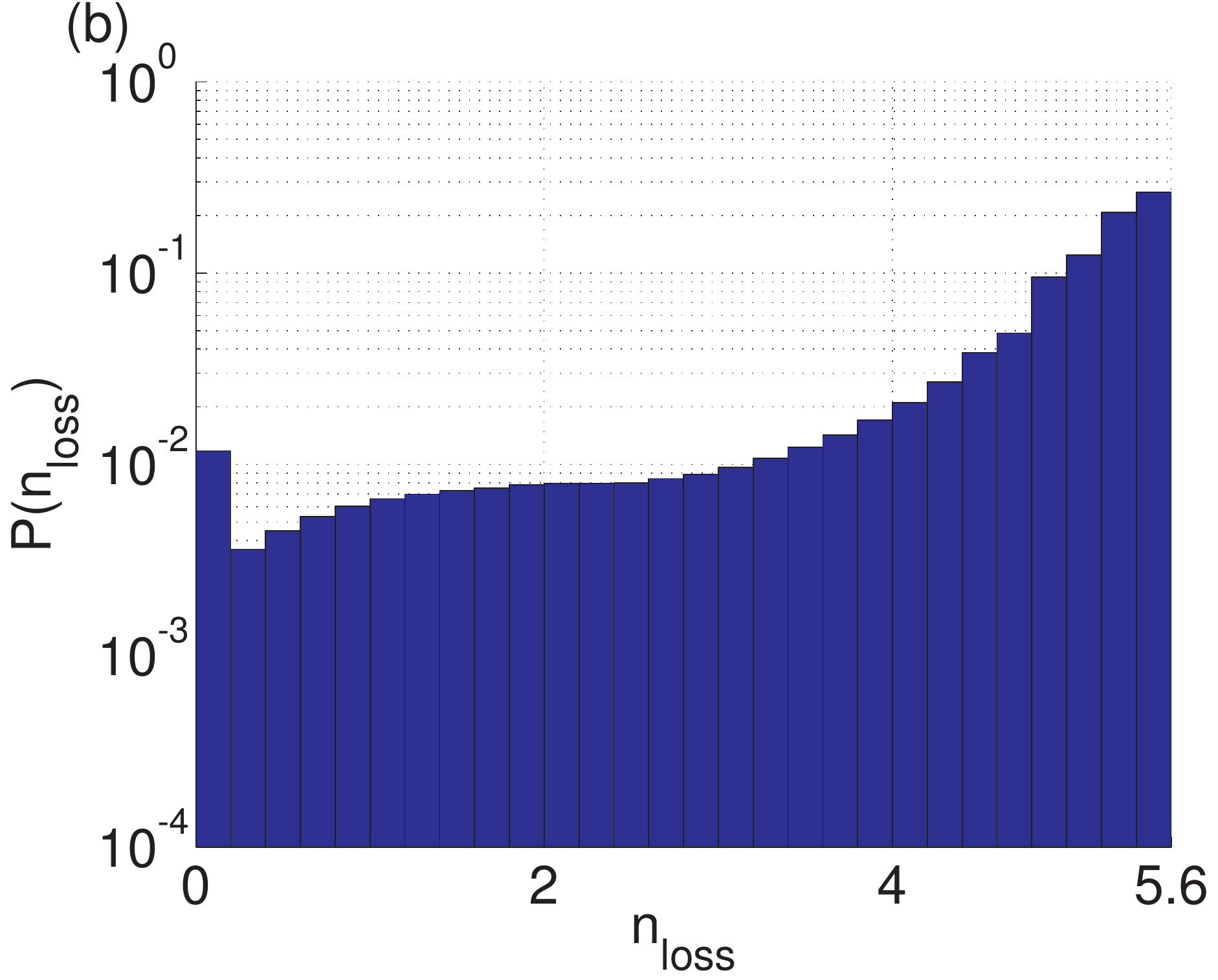}
\end{subfigure}%

\begin{subfigure}{.45\textwidth}
\centering
\includegraphics[scale=0.325]{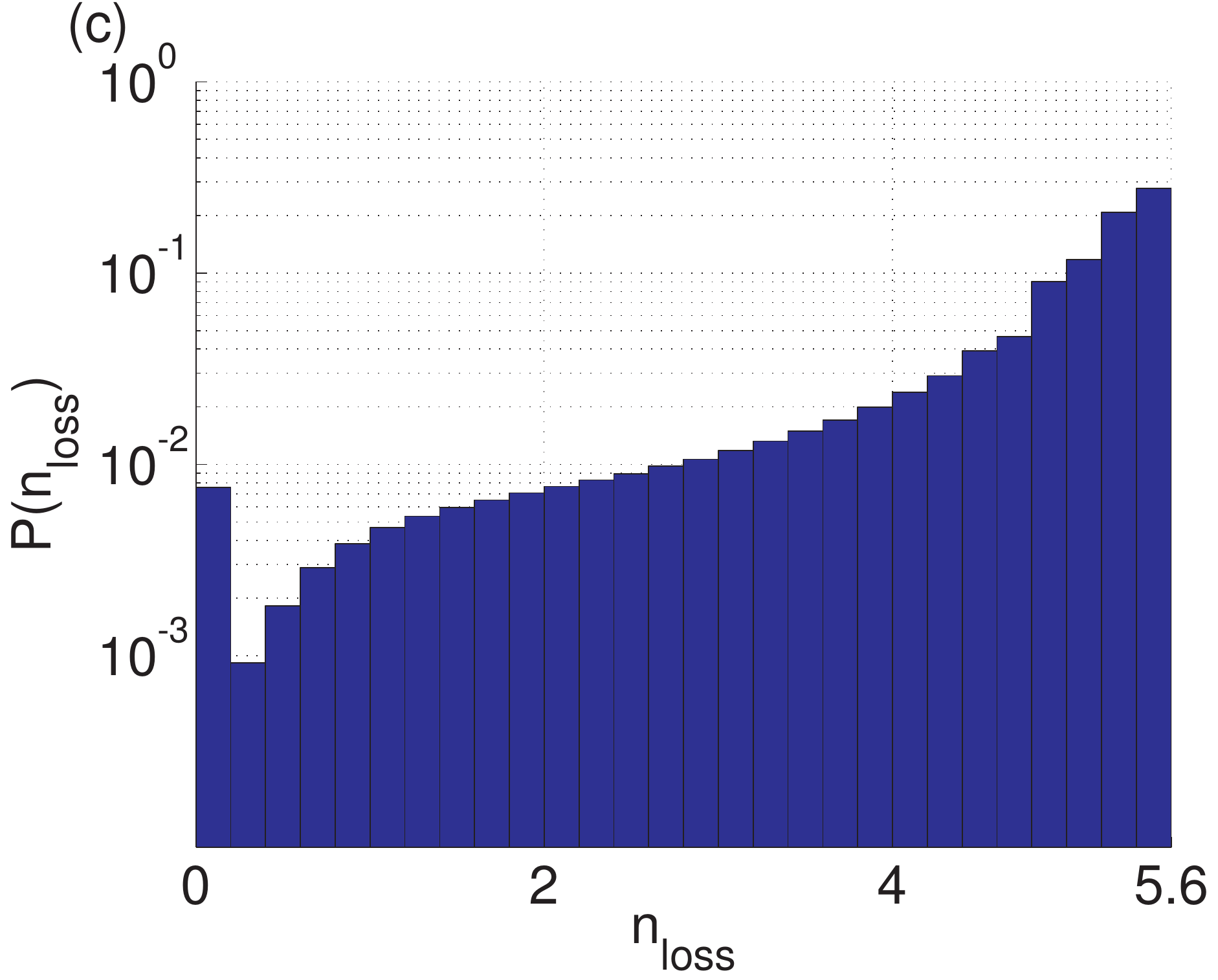}
\end{subfigure}%
\begin{subfigure}{.45\textwidth}
\centering
\includegraphics[scale=0.325]{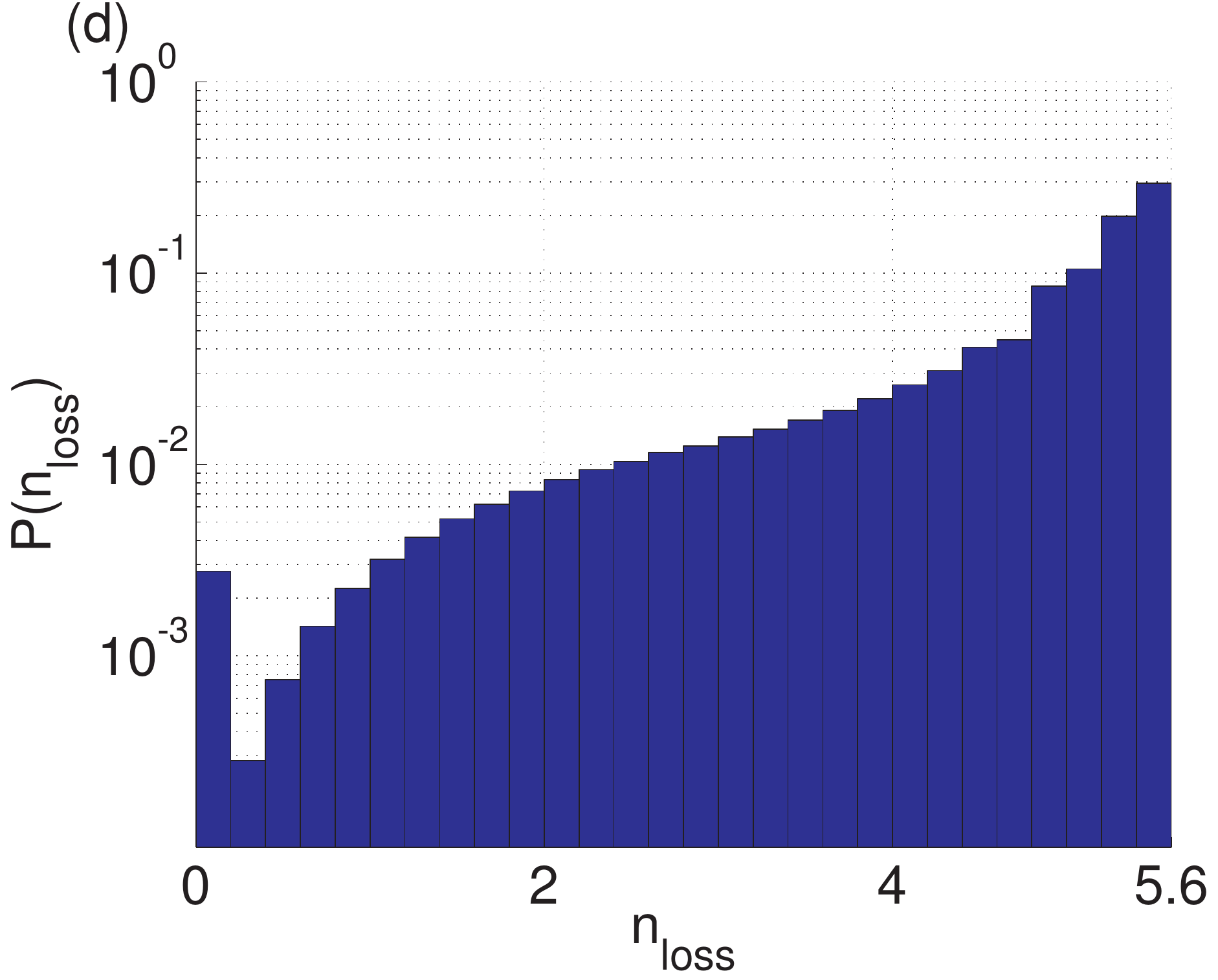}
\end{subfigure}%

        \caption{Probability distribution $P(n_{\rm loss})$ 
         of eruptions with percentage of gas loss $n_{\rm loss}$,
         using data from $10^8$ eruptions with $L=64$ for varying $\delta$:
(a) $\delta=0.1$; (b) $\delta=0.2$; (c) $\delta=0.4$; (d) $\delta=0.6$.}
        \label{fig:Pnlossdelta}
\end{figure*}

 Histograms of the probability density $P(n_{\rm loss})$ 
of events with mean gas-loss $n_{\rm loss}$, 
for varying $\delta$, are shown in Figure~\ref{fig:Pnlossdelta}.
Compared to the case $\delta=0$ shown in Figure~\ref{fig:Pnloss},
for $\delta=0.1$ the exponential relationship 
(i.e.\ the straight line section on the log-linear plot)
is now restricted to $n_{\rm loss}\gtrsim 3.5\%$.
For smaller $n_{\rm loss}$, the probability density function is now much flatter;
i.e.\ high explosivity events with small gas loss 
are now more common.  
For intermediate $\delta$ ($\delta \simeq 0.2$), 
a pronounced peak develops at $n_{\rm loss}\simeq 0$
(i.e.\ the most explosive events are now comparatively common).
For higher $\delta$ ($\delta > 0.2$),
while this peak of ultra-explosive events remains relatively prominent,
the probability of explosive events more generally decreases again.

\par Once magma has moved into a fractured cell 
and the stress level is reset to zero, 
that batch of magma must wait for the cell to fracture again
before being able to move onwards towards an eruption.
If this process takes a particular time on average, 
in the {earlier} model {($\delta=0$)},
then that time will decrease in the new model,
by an amount depending upon $\delta$.
Note that there is a nonlinear effect at work here.
The more cells contain magma, the greater the net strain rate, 
and the more frequent the occurrence of fracture events;  
the more frequent the occurrence of fracture events, 
the more magma is able to enter the system. This magma does not reside in the system for as long however;
the increased seismicity allows the magma to be transported more
efficiently through the system.
This mechanism also contributes to the increased likelihood of large eruptions.
Since clusters of cells containing magma all have the increased strain rate, 
they are all more likely to be involved in further fracture events,
one of which may ultimately cause an eruption.

\par {Note that the nonlinear nature of the link 
between magma and stress fields,
and the dependence of the behaviour on the statistical distributions
of both these fields,
means that there is not a critical value of $\delta$
at which the statistical behaviour of the system changes
(and which would therefore clearly mark different regimes).
Rather, as is clear from Table~\ref{tab:stats},
the effects of increasing $\delta$ can act in differing ways,
resulting in non-monotonic behaviour,
and in different measures achieving extreme values 
at different values of $\delta$.
%
Nevertheless, the behaviour with increasing $\delta$
can be understood in terms of the dynamics 
of magma movement with the model,
as discussed below.}

\par
\begin{figure*}[tbp]
        \centering

\begin{subfigure}{.45\textwidth}
\centering
\includegraphics[scale=0.325]{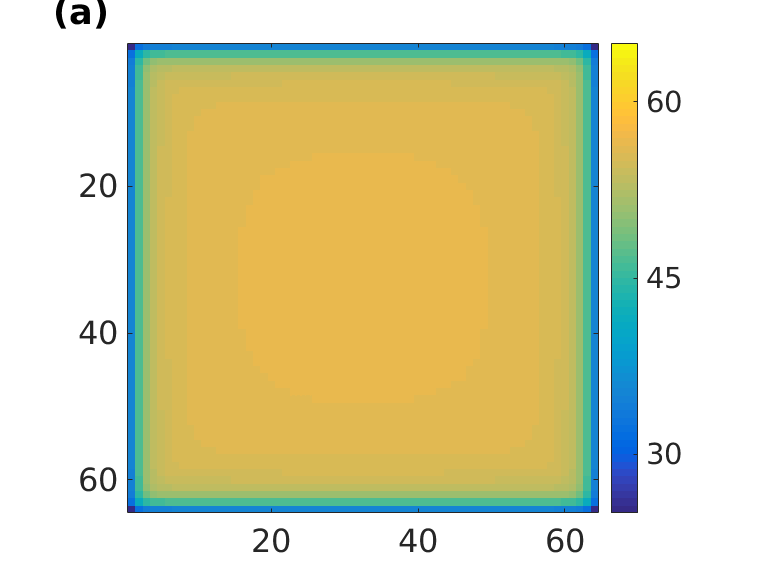}
\end{subfigure}%
\begin{subfigure}{.45\textwidth}
\centering
\includegraphics[scale=0.325]{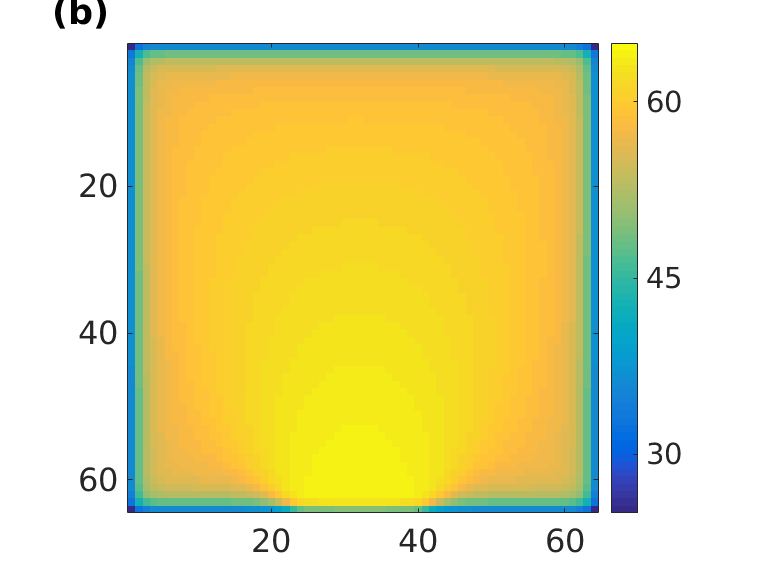}
\end{subfigure}%

        \caption{Frequency distributions of cells involved in fractures, for: 
         (a) $\delta=0$; (b) $\delta=0.4$.
         The numbers on the colour bar correspond to millions of fracture
         occurrences, from a simulation with $L=64$ and $10^8$ eruptions.}
        \label{fig:cells_fracs}
\end{figure*}

 To clarify the dynamics for $\delta>0$,
we have analysed the frequency of occurrence of events within each cell.
Figure~\ref{fig:cells_fracs} 
shows the frequency distribution of fracture events
for $\delta=0$ and $\delta=0.4$.
With $\delta=0$ the distribution is homogeneous,
except for a decrease in the number of events near the edges of the grid,
where the effect of the boundaries 
leads to cells receiving less stress via redistributions 
from events at neighbouring cells.
This homogeneity is to be expected, given the constant applied stress, 
and lack of feedback from magma occupation.
In contrast, with $\delta>0$, fracture events become increasingly 
concentrated in the lower central section of the grid, 
immediately above the magma reservoir. Again, this should be expected, 
since the mechanism for magma entering the chamber 
requires magma to travel through this part of the grid;
and since magma occupation now leads to increased strain rates, 
this must ultimately lead to more fracture events there.
(Although not shown here, the frequency distribution of magma occupation supports this conclusion.) 
In physical terms, the country rock in this region is repeatedly failing
under the net increased stress, and so is experiencing static fatigue.
The part of the grid experiencing repeated failures will clearly depend 
on the width of the opening connecting the magma chamber to the reservoir
(here the central quarter of the grid).
\citet{Piegari2011} stated that the width of the reservoir opening 
did not alter the statistical properties of the {earlier} model,
but the addition of magma-induced stress might well change this conclusion.
This has not yet been explored.

\begin{figure*}[tbp]
        \centering

\begin{subfigure}{.45\textwidth}
\centering
\includegraphics[scale=0.35]{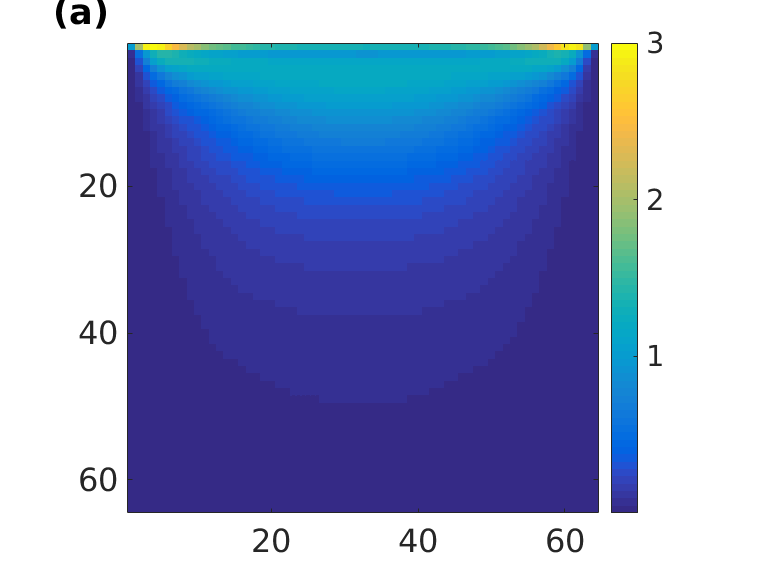}
\end{subfigure}%
\begin{subfigure}{.45\textwidth}
\centering
\includegraphics[scale=0.35]{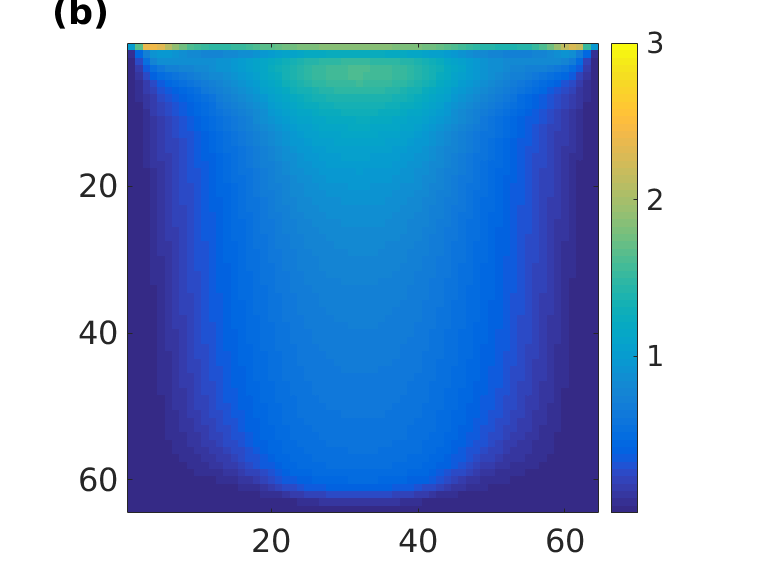}
\end{subfigure}%

\begin{subfigure}{.45\textwidth}
\centering
\includegraphics[scale=0.35]{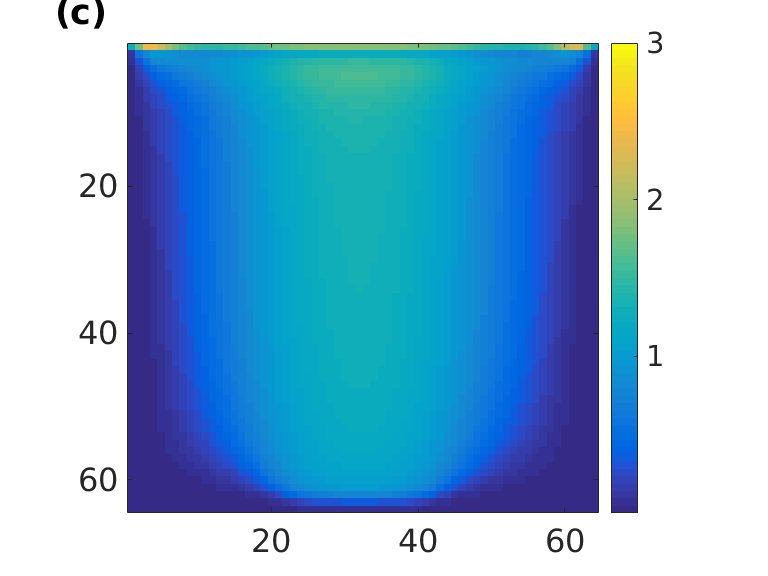}
\end{subfigure}%
\begin{subfigure}{.45\textwidth}
\centering
\includegraphics[scale=0.35]{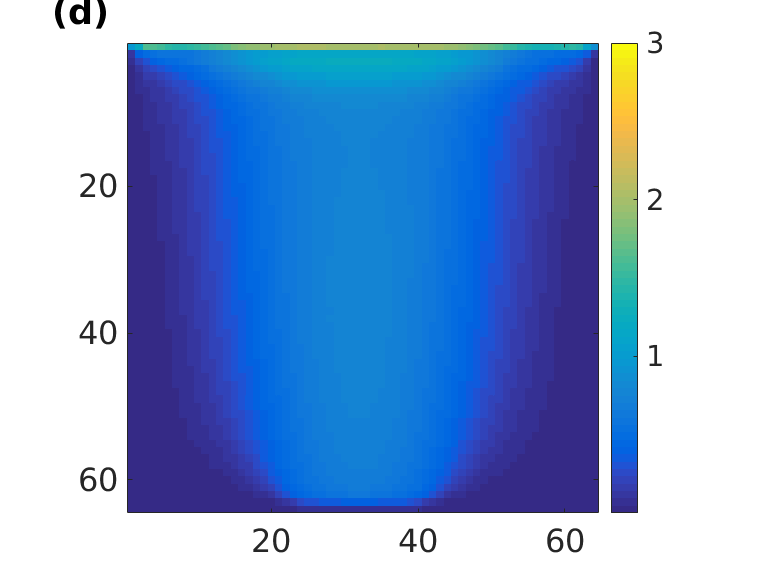}
\end{subfigure}%

        \caption{Frequency distributions of cells involved in eruptions, for:
         (a) $\delta=0$; (b) $\delta=0.1$;
         (c) $\delta=0.2$; (d) $\delta=0.4$.
         The numbers on the colour bar correspond to millions of eruption
         occurrences, from a simulation with $L=64$ and $10^8$ eruptions.}
        \label{fig:cells_erup}
\end{figure*}

\par The frequency distribution of cells involved in eruptions
is shown in Figure~\ref{fig:cells_erup}.
From the panels with $\delta>0$, 
it is clear that cells at greater depths take part in eruptions 
more frequently than for $\delta=0$. 
This is particularly the case within a central column 
above the opening to the magma reservoir. 
The appearance of this central column is a clear signature 
of the presence of magma-induced stresses. 
Eruptions involving deep-lying magma include those large eruptions 
causing the {large-$V$} peak in the $P(V)$ distribution noted above, 
but also smaller, volatile-rich 
(low $n_{\rm loss}$, explosive) eruptions.
In contrast, when $\delta=0$, the cells involved in eruptions 
more frequently come from only the upper regions of the grid, 
and so eruptions more often contain less volatiles (so are more effusive).
For larger $\delta$ ($\delta \gtrsim 0.4$), 
the effectiveness of the magma-induced stress 
in assisting the vertical transport of 
{magma through the central conduit is such that 
the preference towards large, highly-explosive eruptions reverses,
and smaller, shallower events again become relatively more likely.
This is because magma batches can now ascend more `continuously';
i.e.\ via more frequent small steps.
This explains the reduction of the large-$V$ peak 
in the $P(V)$ distribution, observed for $\delta\gtrsim 0.2$ 
in Figure~\ref{fig:PVdelta}.}

\par With regards to Figure~\ref{fig:cells_erup},
it may be worth noting two minor artefacts in the distributions of
near-surface eruptions that arise due to our CA model. 
There are local peaks in the eruption frequency near the edges of our domain,
due to the finite width of our magma chamber.
(This is most pronounced for $\delta=0$, but persists also for $\delta>0$.)
A wider domain  would smooth out this feature.
There is also a local minimum in the eruption frequency in row 2, 
due to a combination of the edge effect in the fracture distribution
(shown in Figure~\ref{fig:cells_fracs}) 
and our magma movement and eruption algorithms.
We do not think these phenomena affect our overall eruption statistics.

\par To summarise, the addition of magma-induced stress has allowed for volcanic activity involving relatively more frequent large, explosive events. 
Within the fracture-network model of magma batch migration, 
this is therefore closer to a more continuous central conduit model,
in which magma does not move too far horizontally 
away from the region above magma reservoir.
Within the classification of \citet{Scandone2009}, 
the eruptions are moving towards a more sustained style.
It is important to appreciate that this feature has not been imposed \citep[along the lines of][who adapted their magma movement algorithm to enforce axial symmetry]{Piegari2012,Piegari2013}. Instead it arises naturally from the increase in stress in regions containing magma. In the VEI index, 
61\% of volcanoes feature central craters \citep{Simkin1993}, so the model 
is arguably now more suited to modelling the majority of volcanoes.


\section{Discussion}
\label{sect:discuss}


We have adopted the model of \cite{Piegari2008,Piegari2011} as the basis for our study, replicating their essential results:
a power law relation between eruption size and frequency;
a stretched exponential distribution of repose times between events,
approaching a pure exponential distribution only for the longest repose times;
and a broadly exponential distribution of eruption volatilities.
This model was then extended to include feedback 
from the magma upon the fracture network, 
via a local magma-related augmentation of the stress field.
This has the effect of creating new peaks 
in the probability distribution $P(V)$,
corresponding to relatively more frequent intermediate and large events,
with the range of power law behaviour being restricted to smaller events.
The mean eruption size consequently increases,
with associated increases in the mean inter-eruption time 
and the likelihood of high explosivity eruptions.

Our new model effectively favours a central `axial' conduit,
as found in many volcano systems \citep[and which otherwise must be artificially imposed, as in][]{Piegari2012}. In this context, it is worth noting that
\cite{Pinel2003} find that the increased stress in the upper crust 
from the weight of edifice volcanoes (e.g.\ stratovolcanoes)
is comparable to tectonic stresses and overpressures within magma 
cavities. Furthermore the highest stress is distributed in such a way that it favours the formation of a central conduit system.
Insofar as our magma-induced stress also  leads to such centralisation,
it might be argued that we are modelling this edifice effect by proxy.

\par There are many additions that might be made to our model. For example, we have not explored the possibility of including a low density surface layer, which would lead to the creation of near-surface dikes and sills, and in the process act as a `cap' on the eruption process \citep{Piegari2012,Piegari2013}. Such a cap leads to a new peak in the probability of long repose time events (on a characteristic timescale controlled by the width  of the low density surface layer), and also leads to fewer `explosive' (high volatile content) events. Given that more explosive events tend to be favoured in our magma-induced stress model, it would be interesting to see how {these two effects would interact}.
We might also consider moving to a 3D model, or perhaps implementing heterogeneous or anisotropic variants 
of the OFC model for the underlying stress field, to reflect the vertical
orientation (with strong pressure gradient) required in the current context. In the existing model, the timescale of magma movement is clearly separated from that of stress-field evolution (the former being effectively instantaneous on the timescale of the latter). We might consider adapting the model to deal with the two timescales in a more continuous way. 

\par While we would argue that we have modelled the local stress enhancement due to the magma in a very natural way (at least within the idealised spirit of the model), there are possible modifications that could be made. For example, we might choose to increase the local stress only around magma batches that moved during the most recent fracture. Alternatively, we might apply some magma-induced stress in cells surrounding those containing magma batches (out to some specified distance), as well as in the magma cells themselves. We could also associate a local stress increase with the occurrence of passive degassing (making the stress proportional to the extent of degassing), to model the occurrence of shallow (long period) earthquakes caused by this process.

\par The addition of magma-induced stress to the CA model opens up new possibilities for studying different styles of volcanism and volcanic-related seismic activity. 
With or without the possible modifications described above, there is much to be gained from comparing our results with those of different models, as well as with real observations. \cite{Sachs2012} consider the possible deviations from power law statistics for earthquakes and volcanoes, and for SOC models of these, focusing on extreme size events that occur more often than extrapolations would suggest (so called `dragon-kings'). In those terms, the peak of large events obtained for $\delta>0$ in our model might be considered dragon-kings.
There are other statistical comparisons that could also be carried out. For example, \cite{Watt2007} considered statistical fits to sequences of Vulcanian explosions, finding support for Weibull and log-logistic fits. \cite{Connor2003} had previously modelled such a sequence from Soufri\`{e}re Hills via a log-logistic distribution, arguing that the log-logistic distribution arises from competing physical processes of gas bubble pressurisation and vesiculation associated with the magma movement. 

{More detailed study of the statistics of variant CA systems ---
incorporating the differing forms of feedback from magma to stress fields 
discussed above (i.e.\ differing forms of local stress enhancement) ---
would allow us to investigate the mechanisms behind such 
statistical distributions.
The present work constitutes a first step in this direction.}

\vspace*{2.5mm}
\noindent{\bf Acknowledgements}
\vspace*{1.25mm}

We thank Ester Piegari for helpful discussions concerning her calculations,
{and thank the reviewers and editor for their comments, 
which helped us to clarify a number of points.}

\vspace*{2.5mm}
\noindent{\bf Bibliography}
\vspace*{1.25mm}



\end{document}